\documentclass[acmlarge]{acmart}
\AtBeginDocument{%
  \providecommand\BibTeX{{%
    \normalfont B\kern-0.5em{\scshape i\kern-0.25em b}\kern-0.8em\TeX}}}

\setcopyright{acmcopyright}
\copyrightyear{2024}
\acmYear{2024}
\acmDOI{XXXXXXX.XXXXXXX}

\acmConference[Proc. ACM Interact. Mob. Wearable Ubiquitous Technol.]{June 03--05,
  2024}{}

\usepackage{graphicx}
\usepackage{lipsum}
\usepackage{algorithm}
\usepackage{algpseudocode}
\usepackage{amsthm}

\theoremstyle{remark}

\theoremstyle{definition}

\usepackage[absolute]{textpos}
\usepackage{amsmath,amsfonts,amsthm}
\usepackage{newtxmath}
\usepackage{mathtools}
\usepackage{commath}
\usepackage{booktabs}
\usepackage{enumitem}
\usepackage{xspace}
\usepackage{microtype}

\usepackage{subcaption}

\usepackage[normalem]{ulem} % \sout command
\usepackage{xcolor}
\usepackage{todonotes}
\usepackage{url}

\newcommand{\name}{\mbox{\textit{PedHat}}\xspace}

\usepackage[normalem]{ulem}

\usepackage{tikz}

\setcopyright{none}
\renewcommand\footnotetextcopyrightpermission[1]{} % removes footnote with conference information in first column

\begin{document}

% \title[OHA]{Real-Time Pedestrian Road Crossing Prediction Using Commodity Devices}

\title[OHA]{Reliable Heading Tracking for Pedestrian Road Crossing Prediction Using Commodity Devices}


% \keywords{datasets, neural networks, gaze detection, text tagging}

\author{Yucheng Yang}
\email{yang552@wisc.edu}
\affiliation{%
  \institution{University of Wisconsin-Madison}
  \streetaddress{1415 Engineering Dr}
  \city{Madison}
  \state{WI}
  \postcode{53706}
  \country{USA}
}

\author{Jingjie Li}
\email{jingjie.jj@ed.ac.uk}
\affiliation{%
  \institution{University of Edinburgh}
  \department{School of Informatics}
  \streetaddress{10 Crichton Street}
  \city{Edinburgh}
  \postcode{EH8 9AB}
  \country{United Kingdom}
}

\author{Kassem Fawaz}
\email{kfawaz@wisc.edu}
\affiliation{%
  \institution{University of Wisconsin-Madison}
  \streetaddress{1415 Engineering Dr}
  \city{Madison}
  \state{WI}
  \postcode{53706}
  \country{USA}
}

\renewcommand{\shortauthors}{Yang et al.}


\begin{abstract}

Pedestrian heading tracking enables applications in pedestrian navigation, traffic safety, and accessibility. Previous works, using inertial sensor fusion or machine learning, are limited in that they assume the phone is fixed in specific orientations, hindering their generalizability. We propose a new heading tracking algorithm, the Orientation-Heading Alignment (OHA), which leverages a key insight: people tend to carry smartphones in certain ways due to habits, such as swinging them while walking. For each smartphone attitude during this motion, OHA maps the smartphone orientation to the pedestrian heading and learns such mappings efficiently from coarse headings and smartphone orientations. To anchor our algorithm in a practical scenario, we apply OHA to a challenging task: predicting when pedestrians are about to cross the road to improve road user safety. In particular, using 755 hours of walking data collected since 2020 from 60 individuals, we develop a lightweight model that operates in real-time on commodity devices to predict road crossings. Our evaluation shows that OHA achieves 3.4 times smaller heading errors across nine scenarios than existing methods. Furthermore, OHA enables the early and accurate detection of pedestrian crossing behavior, issuing crossing alerts 0.35 seconds, on average, before pedestrians enter the road range.

\end{abstract}
\maketitle

\section{Introduction}

Tracking an individual's heading enables important applications, including navigation, safety, entertainment, and accessibility~\cite{chen2020deep, abughalieh2020predicting, zheng2020heading, kandalan2020techniques}. For instance, accurately tracking pedestrian heading, not limited to using mobile devices, facilitates the prediction of pedestrian behavior on roads~\cite{sighencea2021review, schulz2015controlled, keller2013will, minguez2018pedestrian, zhang2021pedestrian, fang2019intention, kuang2018robust, roy2014smartphone}. Additionally, heading tracking methods based on inertial sensors installed on portable devices provide continuous and practical heading estimates without requiring dedicated devices such as LiDAR or stereo camera systems, garnering significant interest from researchers in both industry and academia~\cite{wu2019survey, feigl2019bidirectional}.

Tracking pedestrian heading involves continuously tracking an individual's facing direction on a 2-D flat plane, typically the horizontal plane of the global coordinate system (GCS). It is important to note that tracking pedestrian heading using mobile or wearable devices differs from tracking the orientation of these devices in GCS, as explored in previous works such as MUSE~\cite{shen2018closing} and $A^3$~\cite{zhou2014use}. For example, a pedestrian could be walking from south to north on a road while swinging a smartphone. In this case, smartphone orientation estimation would indicate the device's dynamic orientation relative to the GCS, commonly represented by Euler angles (roll, pitch, yaw). On the other hand, tracking pedestrian heading should accurately show that the pedestrian is moving from south to north, regardless of how the smartphone is oriented.

Existing approaches to estimating pedestrian heading through IMU (Inertial Measurement Unit) employ a two-stage pipeline: first, they estimate the horizontal plane using gravity or magnetic fields, and then integrate the gyroscope to track relative heading changes~\cite{Manos2018gravity, thio2021relative, deng2015heading}. These approaches hinge on a critical assumption: the phone must remain static relative to the pedestrian body. Once the phone moves, especially during swinging motions, frequent recalculations of the horizontal plane are necessary, introducing intractable errors into the system, as we describe in Section~\ref{sec:oha_accu}. Similarly, researchers have explored tracking pedestrian trajectories using IMU sensors through methods such as dead reckoning~\cite{fan2019improved} or machine learning~\cite{chen2018ionet}. However, these methods suffer from drifts and inaccuracies inherent to on-device sensors, preventing them from accurately tracking fine-grained pedestrian heading or movements over extended periods~\cite{gong2021robust}.

We propose a new heading tracking algorithm, Orientation-Heading Alignment (OHA), which leverages a key insight: people tend to carry smartphones in certain attitudes due to habits, whether swinging them while walking, stashing them in pockets, or placing them in bags. These attitudes or relative orientations, defined as the smartphone's orientation relative to the human body rather than GCS, mainly depend on the user's habits, characteristics, or even clothing. For instance, regardless of which direction a pedestrian faces, they swing the smartphone in their habitual manner. For each smartphone attitude, OHA maps the smartphone orientation to the pedestrian heading. Because the attitudes are relatively stable for each person (e.g., holding a smartphone in the right hand and swinging), it is possible to learn the mappings efficiently from coarse headings and smartphone orientation.
Previous research~\cite{liu2023pedestrian, yang2020wifi, lee2023cnn} has noted a similar insight but adopted a different approach for heading tracking: collecting IMU and accurate heading information for multiple smartphone attitudes and training a machine learning model to predict the heading. However, due to device discrepancies and varying user behaviors, it is not feasible to construct a machine learning model that generalizes to all possible smartphone attitudes.

To anchor our heading estimation algorithm in a practical scenario, we apply OHA to a challenging task: predicting when pedestrians are about to cross the road—an important problem for improving road user safety~\cite{Stewart2023, zhang2021pedestrian, zhang2020prediction}. This task, which requires accurate and timely predictions of pedestrian crossings, is further complicated by the diverse crossing patterns of pedestrians and the complexity of road layouts. Based on the OHA heading, we propose \name, a lightweight, infrastructure-free system that predicts when a pedestrian is about to cross the nearest road and issues crossing alerts. \name incorporates a lightweight model that accepts OHA headings as inputs and operates in real-time on user devices to predict road crossings. We developed this model using data we collected since 2020 from 60 individuals, each contributing two months of traces, covering 755 hours of walking data. We generated coarse-grained labels for these data traces, identifying and eliminating uncertain crossing behavior caused by severe GPS drifts. We integrated this functionality into a lightweight Android app that requires no special hardware, infrastructure, or user involvement. This app autonomously predicts road crossings and alerts nearby road users, making it easier to deploy and adopt than other solutions requiring dedicated infrastructure or special hardware.

We evaluated both OHA and \name in real-world settings, and here are our evaluating highlights:

$\bullet$ We compared OHA with two other baseline methods in outdoor settings: 1) GPS and 2) integrating horizontal components of the gyroscope. We tested them across nine distinct scenarios, which included three walking patterns and three smartphone placements. OHA achieved 3.4 times smaller heading errors across all scenarios than the gyroscope integration method. Moreover, OHA maintains low heading errors, even when participants walked in ``Multiple S-shaped Paths'' while swinging their smartphones. Additionally, OHA demonstrates reliable performance even in conditions of low GPS accuracy or unavailable GPS bearing.

%%%%%%%%%%%% COMBINE THE BELOW TWO POINTS INTO ONE BULLET POINT 
$\bullet$  We conducted a user study with 25 additional participants in a semi-controlled real-world environment to evaluate \name in two key aspects: the accuracy in identifying road crossing events and the promptness of the crossing alerts. Utilizing fine-grained labels from two annotators, we found that \name achieved a precision of 86.9\% and a recall of 93.6\% in identifying road crossing events. Our evaluation also shows that \name issues a crossing alert on average 0.35 seconds before a pedestrian crosses the road edge, delivering performance comparable to camera-based methods that identify pedestrian pose to predict crossings in Line-of-Sight situations~\cite{fang2019intention, zhang2021pedestrian, jeong2016early, keller2013will, xu2011detection}.

$\bullet$ We measured the performance overhead of \name, including OHA on multiple Android devices. We found the average execution time for each prediction is around 10 ms, and the average battery consumption over 30 minutes is about 95 mAh in idle mode and about 352 mAh in active mode.

\section{Preliminaries}

Before describing pedestrian heading tracking, we provide the necessary background on smartphone coordinate systems and how to transform one coordinate system into another. 

\subsection{Coordinate Systems}
Smartphones report their dynamic state in three axes relative to the screen, which we call the local coordinate system (LCS). In this paper, we follow the convention of the Android platform when describing the three axes. The X-axis is horizontal and points to the right of the screen, the Y-axis is vertical and points to the screen top, and the Z-axis points outward from the screen face.
The global coordinate system (GCS) is a three-dimensional reference frame used to locate objects on the Earth. We use the following convention to represent axes in the GCS: the X-axis aligns with the east direction (E), the Y-axis aligns with the magnetic north direction (N), and the Z-axis points opposite to the direction of gravity (G). Due to differences in coordinate systems, raw IMU sensor measurements from a smartphone describe the phone's dynamic state in the LCS rather than the GCS. 
% Additional information is needed to track the phone motion in GCS from IMU sensor readings.

\begin{figure}
    \centering
    \includegraphics[width = 0.6\linewidth]{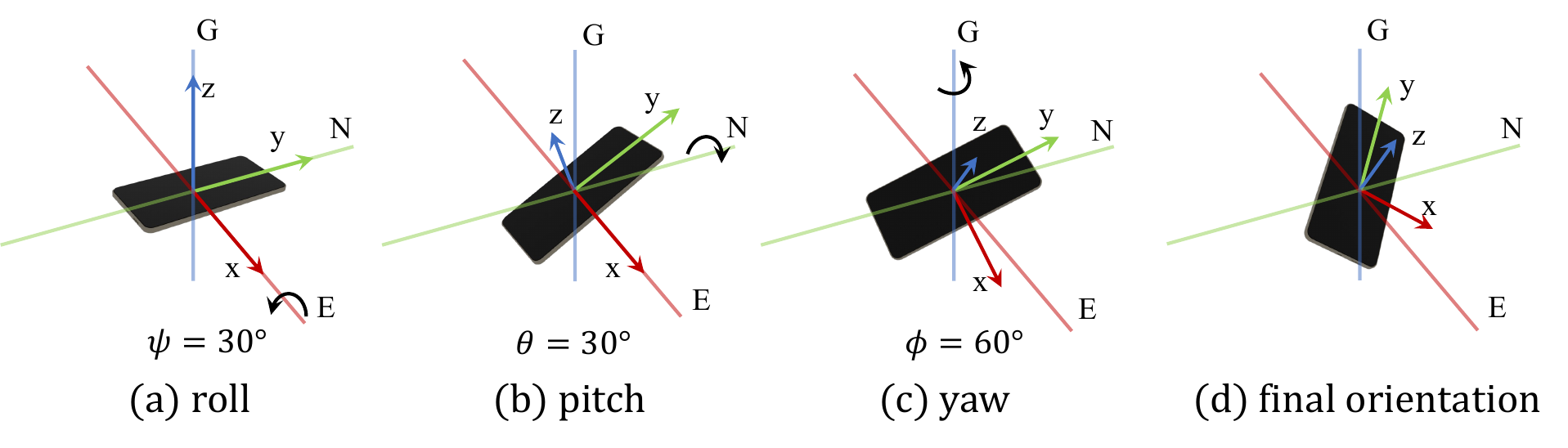}
    \caption{Smartphone's LCS initially aligned with GCS, (a) roll first by $\psi$, (b) pitch second by $\theta$, and (c) yaw third by $\phi$, to achieve (d) last orientation.}
    \label{fig:euler_angles}
    \Description{}
\end{figure}

\subsection{Motion Tracking}
Tracking the phone motion in GCS requires describing its movement along six degrees of freedom (6DOF), including translational and rotational movements. The translational degrees of freedom represent movement along the $X$, $Y$, and $Z$ axes, while the rotational degrees of freedom correspond to rotation around these axes, often referred to as Euler angles (roll, pitch, and yaw). The phone's orientation relative to the Earth allows for transforming IMU sensor readings from LCS to GCS.  

Euler angles provide an intuitive way to describe an object's orientation in 3D space.
They refer to a set of three angles, $\psi, \theta, \phi$, that define the orientation of a rigid body in three-dimensional space relative to a fixed reference frame, in our case, the GCS.
Specifically, they are the rotation angles required to transform one coordinate system to another following a specific rotating order. 
In our setting, Euler angles describe how we rotate a phone that aligns with the GCS ($x,y,z$ aligns with $E,N,G$) to its current orientation following a standardized order. 
For instance, the roll-pitch-yaw convention, also known as $ZYX$ convention, represents a rotating sequence to rotate the GCS into LCS, following a roll ($X$ axis, East) first by $\psi$ degrees, pitch ($Y$ axis, North) second by $\theta$ degrees, and yaw ($Z$ axis, opposite of gravity) last by $\phi$ degrees order.

Figure \ref{fig:euler_angles} illustrates the rotation of a smartphone, initially aligned with the GCS, to its current orientation. Initially, the smartphone's screen coordinate system (LCS) is aligned with the GCS, where $x$ corresponds to the $E$ direction, $y$ to the $N$ direction, and $z$ to the $G$ direction. First, the smartphone rotates around the $E$ direction by $\psi$ (roll) degrees, following the right-hand rule. Then, the smartphone rotates around the $N$ direction by $\theta$ (pitch) degrees. It is important to note that after the first rotation, the $y$ axis is no longer aligned with the $N$ direction. Finally, the smartphone rotates around the $G$ direction by $\phi$ (yaw) degrees, resulting in its final orientation.

\subsection{Transformations}
Orientation information can also be represented as rotation matrices.
Rotation matrices are $3 \times 3$ matrices that transform a coordinate or a vector from one frame (e.g., LCS) to another (e.g., GCS). 

A rotation matrix can be expressed using the Euler angles. In the roll-pitch-yaw convention, there exists a one-to-one correspondence between Euler angles and rotation matrices under restricted domain ranges for the roll, pitch, and yaw angles. In such a case, the rotation matrix can be written in the following form: 
\begin{equation}\label{eq:rpy}
    R_{rpy}(\phi,\theta,\psi) = R_Z(\phi) R_Y(\theta) R_X(\psi),
\end{equation}

where $\phi$, $\theta$, and $\psi$ represent the roll, pitch, and yaw, respectively. Each rotation matrix $R_X$, $R_Y$, and $R_Z$ represents pure rotation around that axis\cite{wiki:rotation}; in GCS, axes $X, Y, Z$ correspond to $E, N, G$. For example, $R_N(45^\circ)$ represents rotating the phone around the North direction for 45 degrees following the right-handed rule. Equation \eqref{eq:rpy} indicates that any orientation can be decomposed into three rotation matrices, i.e., three pure rotations around the $E$, $N$, and $G$ axes one by one. From this decomposition, we can compute $\phi,\theta,\psi$, which allows \name to estimate the user's heading, as will be evident later.

\section{Part I: Pedestrian Heading Tracking}\label{sec:rt_heading}

\begin{figure}
    \centering
    \includegraphics[width =0.4\linewidth]{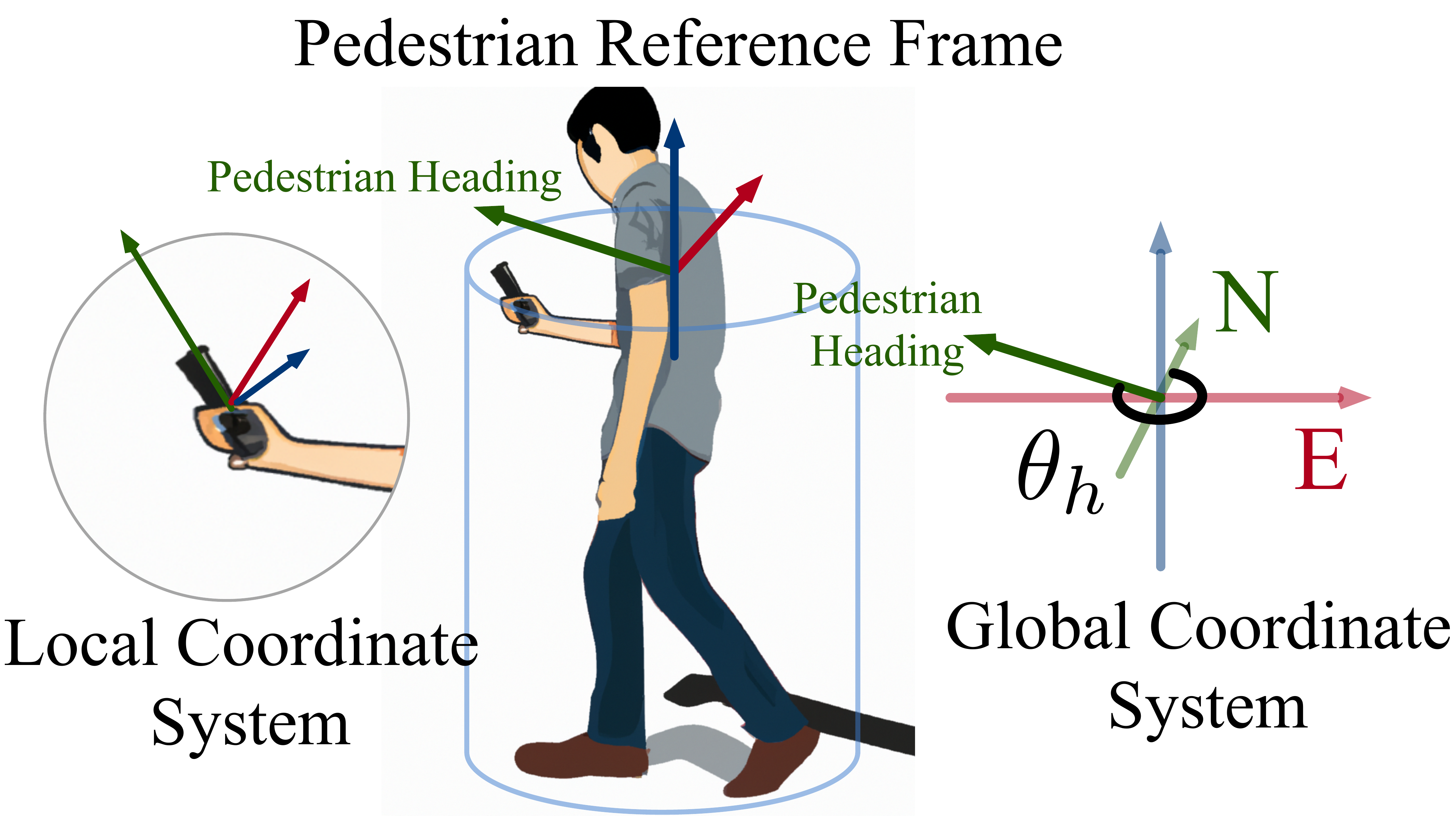}
    \caption{Pedestrian heading $\theta_h$ represents angle between PRF and GCS. A smartphone in front of the pedestrian in LCS. This image was created with the assistance of DALL$\cdot$E 2.}
    \label{fig:PRF}
    \Description{}
\end{figure}

A pedestrian's heading refers to the direction in which a pedestrian is facing projected onto a 2D horizontal plane. Essentially, we ignore any tilt of the body upwards or downwards and focus on the movement along a flat surface. 
% While a pedestrian is walking, we assume the pedestrian heading is the same as their walking direction. When the pedestrian is stationary, the pedestrian heading changes as the pedestrian turns their body. 
Figure~\ref{fig:PRF} illustrates an example of pedestrian heading, and the pedestrian reference frame (PRF) along with LCS and GCS. In the PRF, the Y-axis aligns with the pedestrian heading, and the Z-axis points opposite to the direction of gravity. The pedestrian can turn to change direction and move forward along the Y-axis.

\subsection{Limitations of Current Heading Tracking Methods}

There exist many methods for determining pedestrian heading, ranging from those yielding low to high precision. In outdoor pedestrian tracking, a low-precision pedestrian heading can be derived from GPS bearing, which is the direction computed from two consecutive GPS points. In indoor tracking, low-precision pedestrian heading can similarly be obtained from past trajectory points using methods such as Wi-Fi localization and Map-Matching~\cite{project-osrm}. However, such methods cannot correctly indicate the pedestrian heading when the pedestrian turns without moving. For high-precision pedestrian heading tracking, previous researchers proposed various methods including: 1) utilizing sensors mounted on the human body, such as on foot, chest, and head~\cite{Wang2019pedestrian, zheng2020heading}, 2) camera-based head pose estimation~\cite{perdana2021early,fang2019intention, zhang2021pedestrian}, 3) IMU sensor fusion-based heading tracking~\cite{Manos2018gravity, pei2018optimal, thio2021relative, deng2015heading, yang2021robust}, and 4) machine-learning based heading prediction~\cite{liu2023pedestrian, yang2020wifi, lee2023cnn}. The first two methods require additional hardware for tracking, while the latter two can estimate pedestrian headings using smartphones carried by users. 

Traditional IMU sensor fusion-based heading tracking techniques typically assume that the smartphone is fixed to the human body. These methods generally involve two-stage pipeline: first, they compute a horizontal plane using gravity, the geomagnetic field, or through smartphone orientation estimation. Then, they integrate the horizontal component of the gyroscope to compute the heading changes. An accurate initial heading and a fixed smartphone attitude are often vital for this method. Alternatively, in the second phase, they track the acceleration direction and compare with the geomagnetic north to estimate the heading direction.

With a fixed smartphone attitude, computing pedestrian heading becomes straightforward. However, heading tracking becomes more challenging when the user adopts multiple smartphone attitudes. Other researchers have explored machine-learning-based methods to address this issue~\cite{liu2023pedestrian, lee2023cnn}. They begin by collecting accurate heading data and IMU sensor data across multiple phone attitudes and then train a model to understand the correlation between pedestrian heading and IMU sensor data. However, this approach may fail if the model encounters a new attitude. Overall, these two methods either do not offer a universal tracking solution for phone attitude changes or are effective only under specific phone attitudes.

\subsection{OHA Overview}\label{sec:oha}
Orientation-Heading Alignment (OHA) algorithm designs improvements to pedestrian heading tracking against multiple smartphone attitudes. The key observations is that people tend to carry phones in certain ways or attitudes due to habits, whether swinging them while walking, stashing them in pockets, or placing them in bags. Given the device discrepancy and user behaviors, while it’s not feasible to design a system that can anticipate every potential phone attitude for all users, OHA leverages this opportunity to learn the smartphone's attitudes, referred to as the ``relative orientations'' to human body in the follows, for one user during movement.

\begin{figure}
    \centering
    \includegraphics[width=\linewidth]{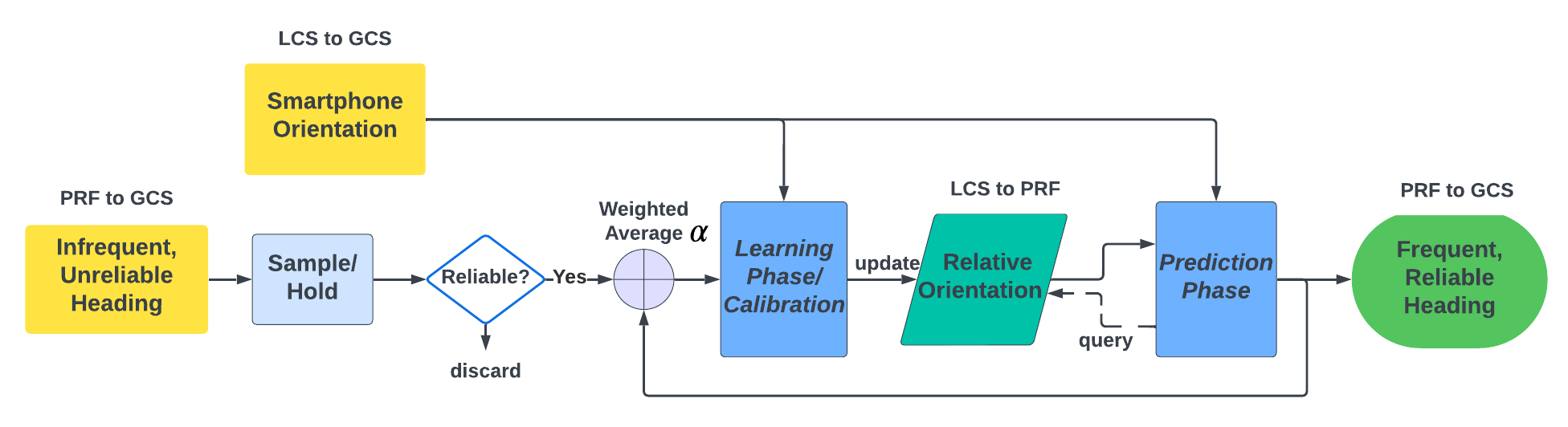}
    \caption{\textbf{OHA processing pipeline}: smartphone orientation data, along with unreliable and low-frequency headings (e.g. 1Hz), are processed in two phases to compute reliable and high-frequency headings (e.g., 50Hz). A weighted average is employed to merge unreliable headings with drift errors, and high-frequency headings with possible cumulative errors. \textbf{Learning Phase}: The weighted average heading and smartphone orientation data are combined to compute and calibrate the smartphone's relative orientation at current timestamp. \textbf{Prediction Phase}: High-frequency smartphone orientation data and the smartphone's current relative orientation are combined to compute a reliable and high-frequency heading.}
    \Description{}
    \label{fig:oha}
\end{figure}

Figure~\ref{fig:oha} illustrates the OHA heading estimation pipeline, which converts infrequent and unreliable headings into frequent and reliable ones. Essentially, OHA learns the relative orientations between smartphones and the human body (from LCS to PRF), and then combines them with smartphone orientation (from LCS to GCS) to estimate a precise pedestrian heading (from PRF to GCS). Initially, OHA lacks knowledge of any relative orientation matrices and utilizes unreliable headings to learn these matrices. As the user moves, OHA continuously calibrates the corresponding relative orientation matrices in the Learning Phase, using a weighted average of both low-frequency headings and high-frequency headings. Additionally, in the Prediction Phase, frequent updates on smartphone orientation allow for the computation of real-time and reliable pedestrian headings. It should be noted that due to our choices for PRF and GCS, the relative orientation matrices and smartphone orientation matrices share the same roll-pitch pair, which we use to index the relative orientation matrices. Therefore, for each smartphone orientation update, OHA queries the corresponding relative orientation matrix and compute a pedestrian heading. Consequently, OHA can produce precise and frequent pedestrian headings for multiple relative orientations, even for motions such as swinging.

\subsection{Formal Analysis}

Here we provide a mathematical proof for OHA. OHA aims to predict the pedestrian heading $\theta_{h,t}$ in real time for time $t$. Note that PRF is the same as rotating GCS around the Z axis by $\theta_{h,t}$ in the clockwise direction, represented by rotation matrix $R_{h,t} = R_Z(-\theta_{h,t})$. In the rotation matrix, rotating in the clockwise direction is negative. 

We assume that at time instant $t$, $O_t = R_{rpy}(\phi,\theta,\psi)$ represents the smartphone orientation, representing rotations from GCS to LCS. $R_{LP, t}$ describes the current relative orientation, representing rotations from \textbf{L}CS to \textbf{P}RF. Alternatively, this rotation $R_{LP, t}$ can be completed through two steps: (1) rotate from LCS to align with GCS $O_t^{-1}$, and (2) rotate from GCS to align with PRF $R_{h, t}$. Therefore, we have:
\begin{align}\label{eq:oha}
\begin{split}
R_{LP,t}^{-1} &= R_{h,t}^{-1} \times O_t = R_Z(\theta_{h,t}) \times R_Z(\phi_t) \times R_Y(\theta_t) \times R_X(\psi_t) \\
&= R_Z(\theta_{h,t} + \phi_t) \times R_Y(\theta_t) \times R_X(\psi_t)
\end{split}
\end{align}

Note that smartphone orientation $O_t$ and relative orientation $R_{LP, t}$ share same pitch ($\theta_t$) and roll ($\psi_t$) values. Therefore, we can index the relative orientations using the roll-pitch pair, and query the corresponding relative orientations using the roll-pitch pair from smartphone orientation $O_t$ in both Learning and Prediction Phases. In the Learning Phase, we use the weight averaged headings as $\theta_{h,t}$ to learn and calibrate relative orientation $R_{LP, t}$. In the Prediction Phase, we extract heading $\theta_{h,t}$ from smartphone orientation $O_t$ and relative orientation $R_{LP, t}$.

The equation above is also effective when the pedestrian changes the relative orientation of their smartphone. Instead of learning a single relative orientation $R_{LP}$, OHA can learn multiple distinguishable relative orientations, each corresponding to a common attitude where a pedestrian might carry their phone. This includes scenarios such as placement in various pockets, holding the device by one's side, or carrying it in the front. We will discuss the limitations of the OHA algorithm in Section~\ref{sec:oha_limit}.

\subsubsection{Algorithm}

We present the pseudo-code for the OHA algorithm in Algorithm~\ref{alg:oha_alg}. 
OHA processes two input streams: the coarse heading, $\theta_{h,c}$, and smartphone orientation, $O_t$, to produce an output stream of precise heading, $\theta_{h,p}$. 
Here, we briefly discuss a few details. 

\noindent\textbf{Initialization}: OHA initiates a new relative orientation, $R_{LP,t}$, using the current smartphone orientation and a coarse heading. Alternatively, if we assume minimal change in the pedestrian’s heading in the recent past, OHA could utilize the most recent estimated heading for this initialization, which we leave for future works.  

\noindent\textbf{Smartphone Orientation Errors}: Smartphone orientation estimation typically consists of gyroscope integration and opportunistic calibration, and is inevitably prone to cumulative errors. To mitigate the impact of smartphone orientation drift and coarse heading biases, we employ a weighted average function, which works as a complementary filter, to make the estimated relative orientation more robust. 

\noindent\textbf{Quantization}: We query the relative orientation dictionary using the quantized roll-pitch pair rather than the raw roll and pitch angles to increase query hit rate. A low hit rate generally leads to fewer opportunities to calibrate $R_{LP,t}$, thereby making it more susceptible to coarse heading bias errors. Nevertheless, setting overly large quantization factors can also induce estimation inaccuracies. We set the quantization factor to 2.

\begin{algorithm}
\caption{Orientation-Heading Alignment Algorithm}\label{alg:oha_alg}
\begin{algorithmic}
\Procedure{OHA}{coarse heading $\theta_{h,c}$, smartphone orientation $O_t: R_{rpy}(\phi,\theta,\psi)$} 
\While{$O_t$ updates}
    \State Quantize $\theta, \psi$ angles 
    \Comment{Increase query hit rate}
    \If{$R_{LP, t}$ \textbf{not} available}
        \State Compute $R_{LP,t}$ from $\theta_{h,c}$ and $O_t$ for once and skip
        \Comment{Initialize $R_{LP,t}$ }
    \EndIf
    \State Obtain new precise heading $\theta_{h,p}$ from Equation (\ref{eq:oha})
    \Comment{Prediction Phase}
    \If{coarse heading $\theta_{h,c}$ available}
        \State Merge $\theta_{h,p}$ and$\theta_{h,c}$, and calibrate relative orientation $R_{LP, t}$
        \Comment{Learning Phase}
    \EndIf
\EndWhile
\EndProcedure
\end{algorithmic}
\end{algorithm}

\subsection{OHA Evaluation}\label{sec:oha_accu}

We evaluate OHA in an outdoor parking lot using GPS for coarse heading inputs and MUSE\cite{shen2018closing} for smartphone orientation estimation. The selection of input sources may vary depending on the scenario. In our case, MUSE is particularly suitable due to its design optimization for frequent movements and its effective use with minimal ferromagnetic interference in outdoor environment. 

\noindent\textbf{Experiment setup:} We evaluate OHA in three typical smartphone placements: (1) handheld in front side, (2) placed in a trouser pocket, and (3) swung by body side while walking (about 60-degree swing angle). For each smartphone placement, we test three distinct walking patterns: (1) \textbf{S}traight with \textbf{O}ccasional 90-degree \textbf{T}urns (SOT), (2) \textbf{S}tationary \textbf{W}hile \textbf{R}otating to different directions (SWR), and (3) walking along a curvilinear trajectory that resembles \textbf{M}ultiple interconnected ``\textbf{S}''-shaped \textbf{P}aths (MSP). We recruit five participants to walk naturally under these nine movement scenarios without further guidance. Each participant walks 3 minutes for each scenario, 9 minutes continuously for each smartphone placement and in total about 30 minutes.

\noindent\textbf{Baseline Comparison:} We compare OHA’s performance against two other techniques in outdoor settings: (1) GPS bearing, and (2) an Integrated Gyroscope (IG) approach. The IG method first utilizes MUSE-based smartphone orientation to compute the horizontal plane and then integrates the horizontal component of the gyroscope as relative heading changes, similar to previous IMU sensor fusion-based works~\cite{Manos2018gravity, pei2018optimal, thio2021relative, deng2015heading, yang2021robust}. We ensure the IG method has an accurate initial heading. All techniques employ the same IMU and GPS data collected from a Google Pixel 6 smartphone. We set gyroscope and accelerometer sampling frequency at 50 Hz, magnetic field sensor at 10 Hz and GPS updates at 1 Hz. 

\noindent\textbf{Ground Truth:} OHA utilizes another chest-mounted smartphone to measure the ground truth heading. We first securely attach this phone to a participant’s chest using a chest mount strap. We calculate the smartphone’s orientation based on the IMU sensor readings, and subsequently derive the azimuth angle. We perform a one-time initial calibration of the azimuth angle by walking along a straight line from east to west, and use this calibrated azimuth angle as the ground truth heading.

\begin{figure}[ht]
    \centering
    \begin{subfigure}[b]{0.32\textwidth}
        \centering
        \includegraphics[width=\textwidth]{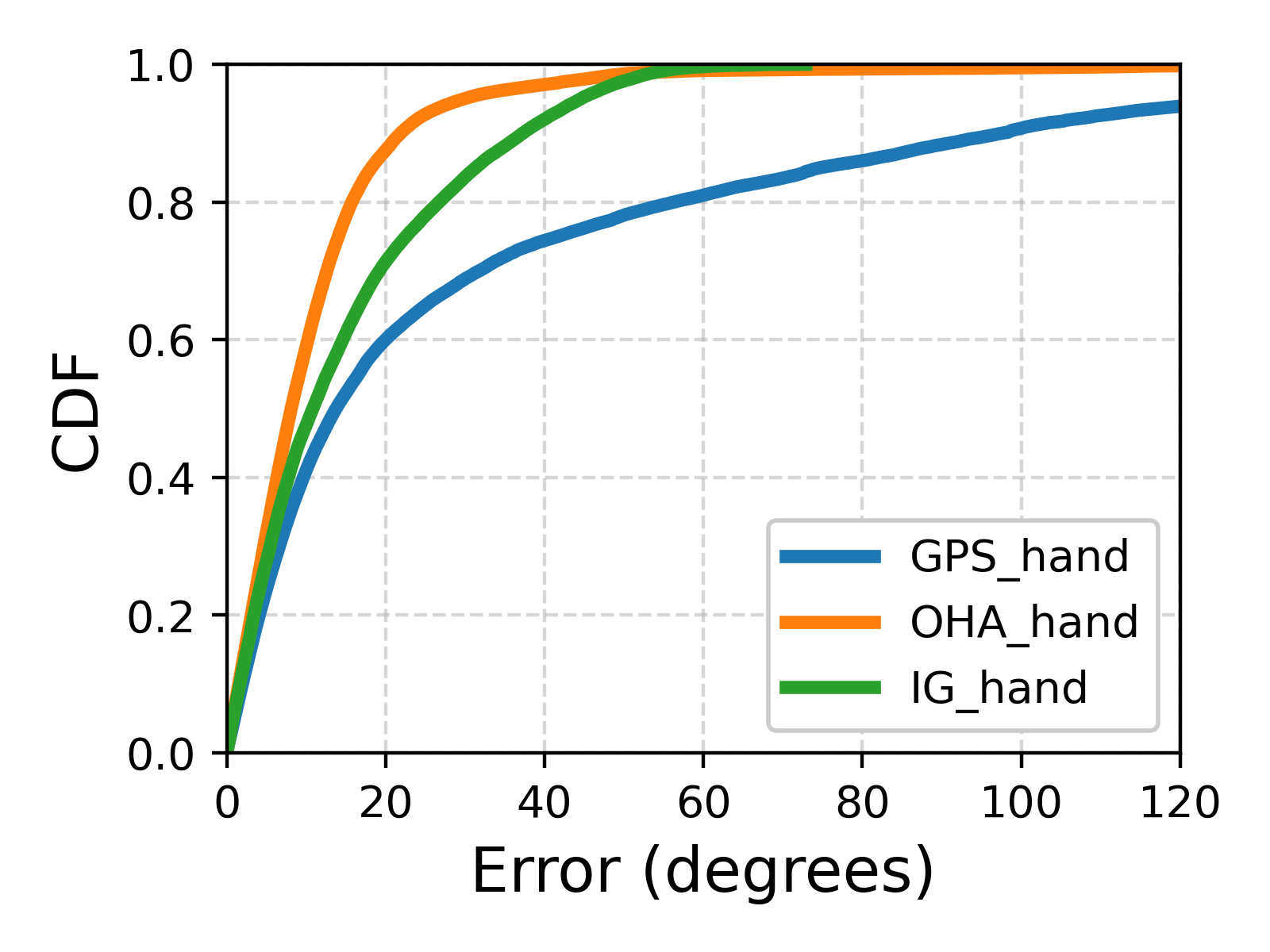}
        \caption{Handhold in front}
        \label{fig:hand_cdf}
    \end{subfigure}
    \hfill % adds horizontal space between figures
    \begin{subfigure}[b]{0.32\textwidth}
        \centering
        \includegraphics[width=\textwidth]{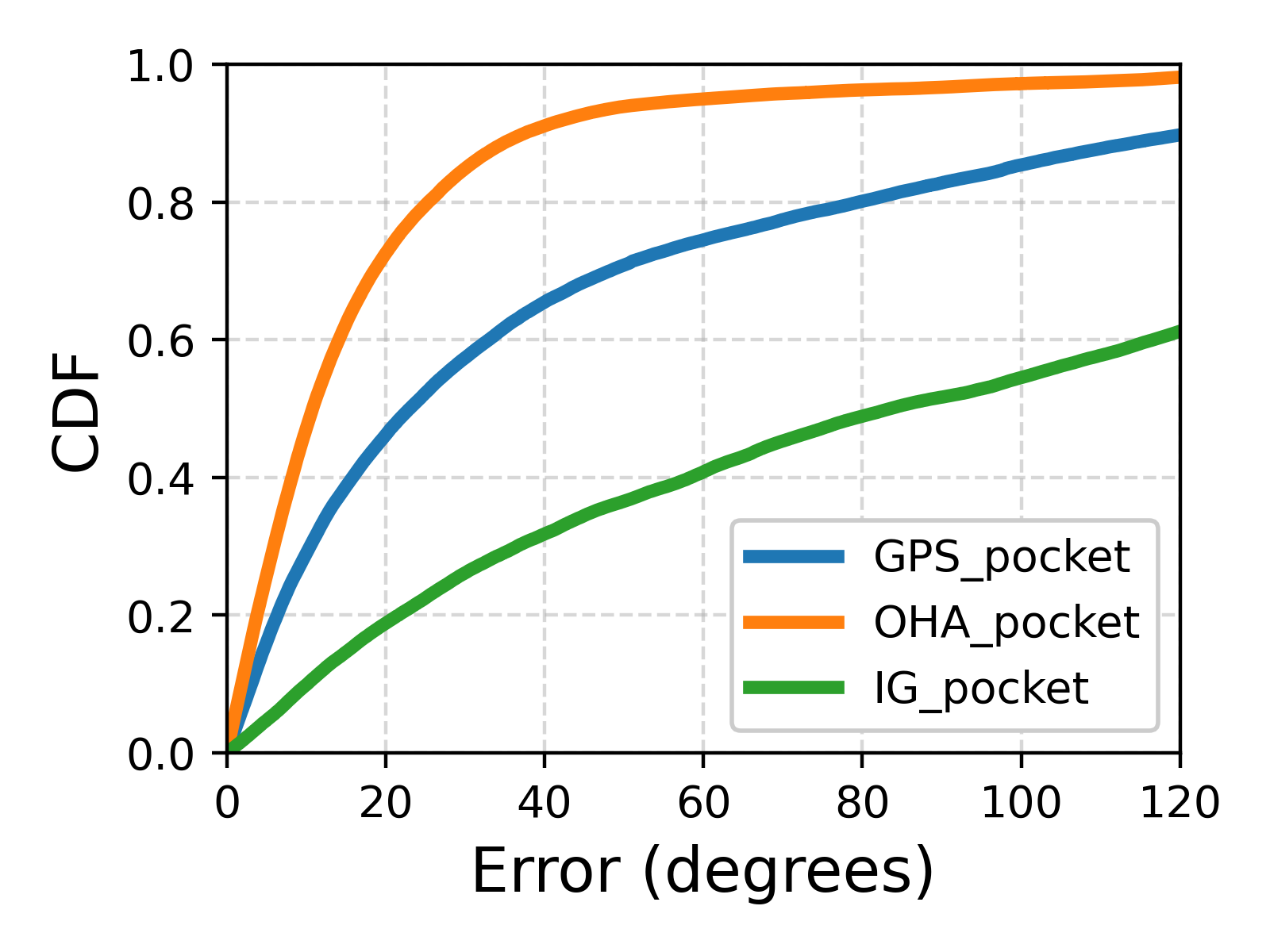}
        \caption{Place in trouser pocket}
        \label{fig:pocket_cdf}
    \end{subfigure}
    \hfill % adds horizontal space between figures
    \begin{subfigure}[b]{0.32\textwidth}
        \centering
        \includegraphics[width=\textwidth]{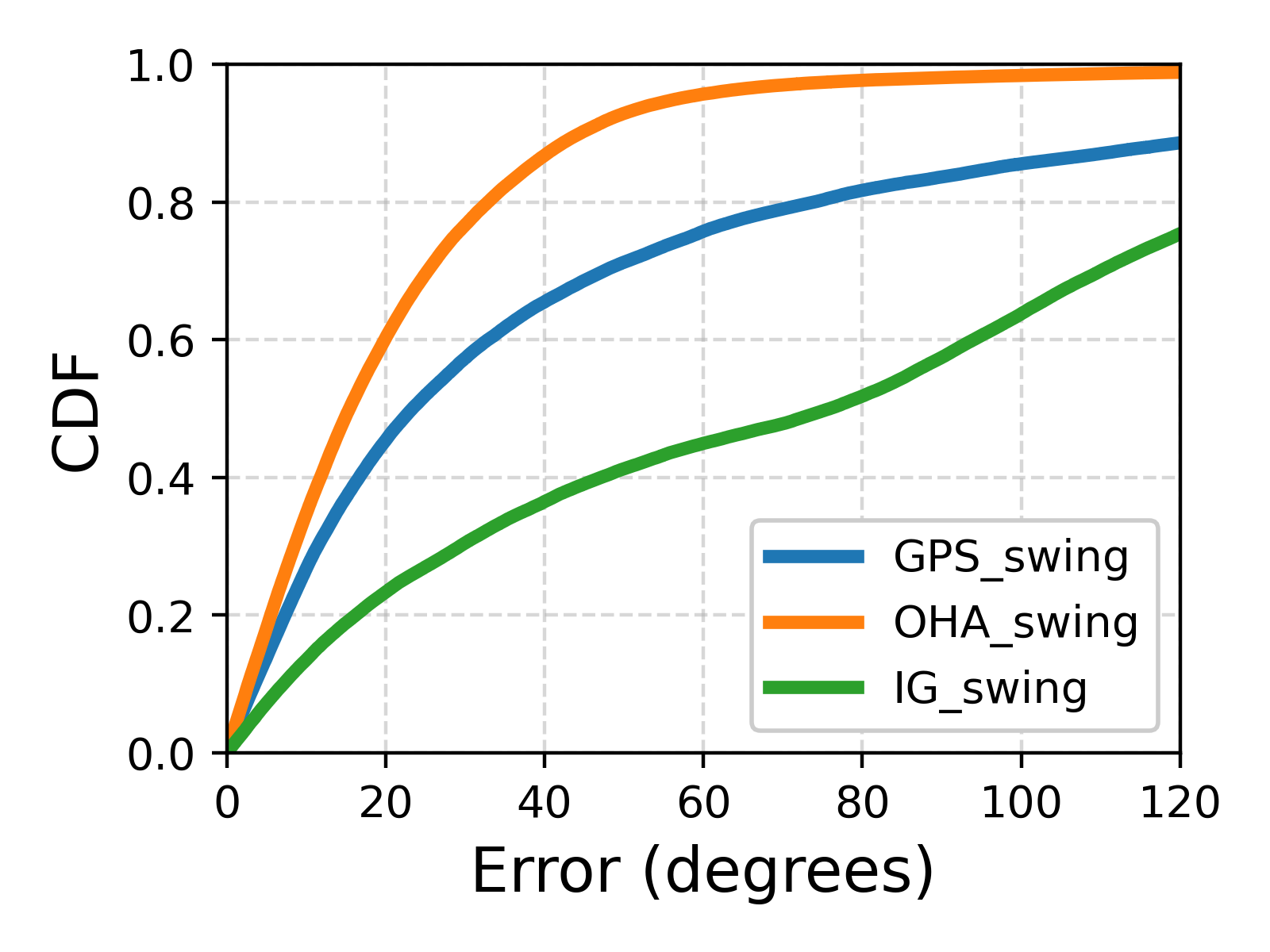}
        \caption{Swing by body side}
        \label{fig:swing_cdf}
    \end{subfigure}
    \caption{Overall heading estimation error CDF for three common smartphone placements: (1) handheld in front, (2) in a trouser pocket, and (3) swinging by the body side with approximately 60-degree swing angles. We compared OHA with GPS bearing and integrated gyroscope (IG) method, based on IMU and GPS data collected from 5 participants over a total of 135 minutes.}
    \label{fig:error_cdf}
    \Description{}
\end{figure}

\begin{figure}[h]
    \centering
    \includegraphics[width=\linewidth]{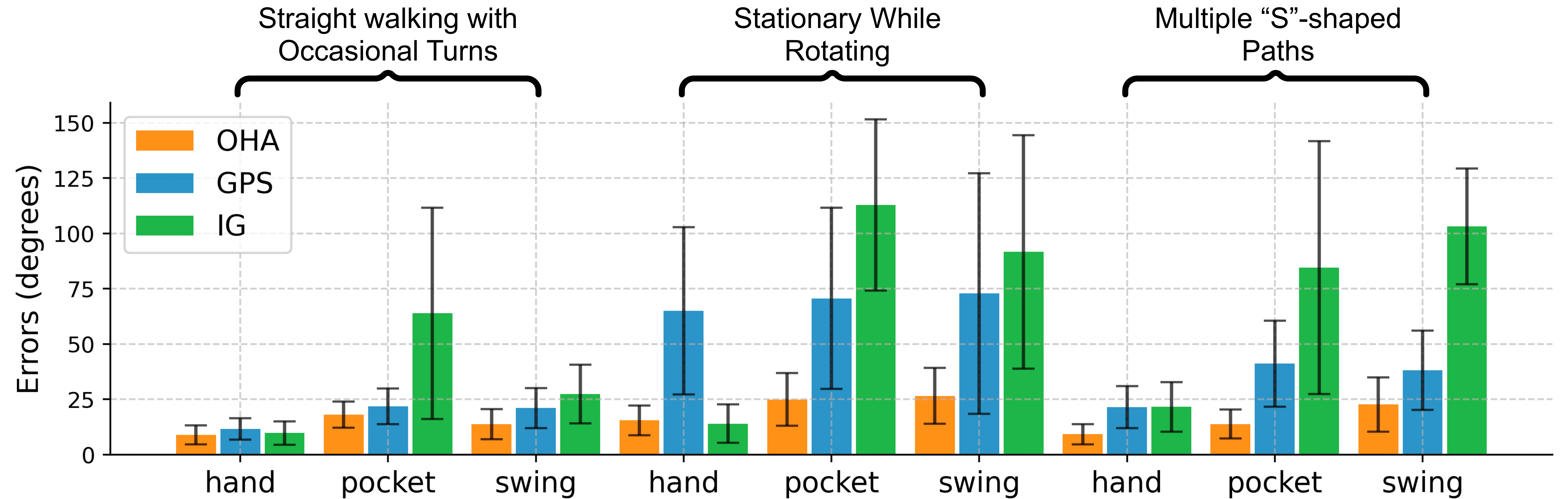}
    \caption{Heading estimation mean and interquartile range errors across 9 movement scenarios from 5 participants. Scenarios combine three smartphone placements: (1) hand, (2) pocket, and (3) swing, and three walking patterns: (1) Straight walking with Occasional Turns (SOT), (2) Stationary While Rotating (SWR), and (3) walking in Multiple ``S''-shaped Paths (MSP).}
    \label{fig:overall_bar}
    \Description{}
\end{figure}
% \todo{the size of the brackets seems weird} solved.
\noindent \textbf{Results:} Figure \ref{fig:error_cdf} compares the overall heading estimation error CDF across three smartphone placements. OHA outperforms the other two techniques consistently in all tested placements. Notably, the performance of the IG method drops significantly when the smartphone is not held in hands. Meanwhile, the GPS heading's performance is almost consistent over different smartphone placements. We observe that when the smartphone is placed in the trouser pocket, it also experiences the swinging motion due to leg movements, similar to that in the swinging case. Therefore the IG method yields higher cumulative errors for both cases. 

Figure \ref{fig:overall_bar} illustrates the mean and interquartile range errors across nine movement scenarios from all five participants. Note that GPS performance drops during ``MSP'' and deteriorates significantly during ``SWR'', whereas OHA consistently maintains low heading errors, underlining its efficacy in conditions where coarse heading inputs are heavily biased. The IG method performs well when the smartphone is held in ``hands'', but its performance is markedly poorer for the other two smartphone placements, especially during ``SWR'' and ``MSP''. In contrast, OHA demonstrates robust accuracy and achieves the lowest error rate in seven out of nine scenarios, except when the smartphone is held in the front during ``SOT'' and ``SWR''. Under these two conditions, where minimal smartphone swinging occurs, the IG method displays similar performance with OHA. Overall, OHA maintains consistent accuracy across different participant and scenarios, outperforming the IG method with an average of 3.4 times smaller errors.

\begin{figure}[ht]
    \centering
    % First figure: Participant Trace
    \begin{subfigure}[b]{0.78\linewidth}
        \centering
        \includegraphics[width=\linewidth]{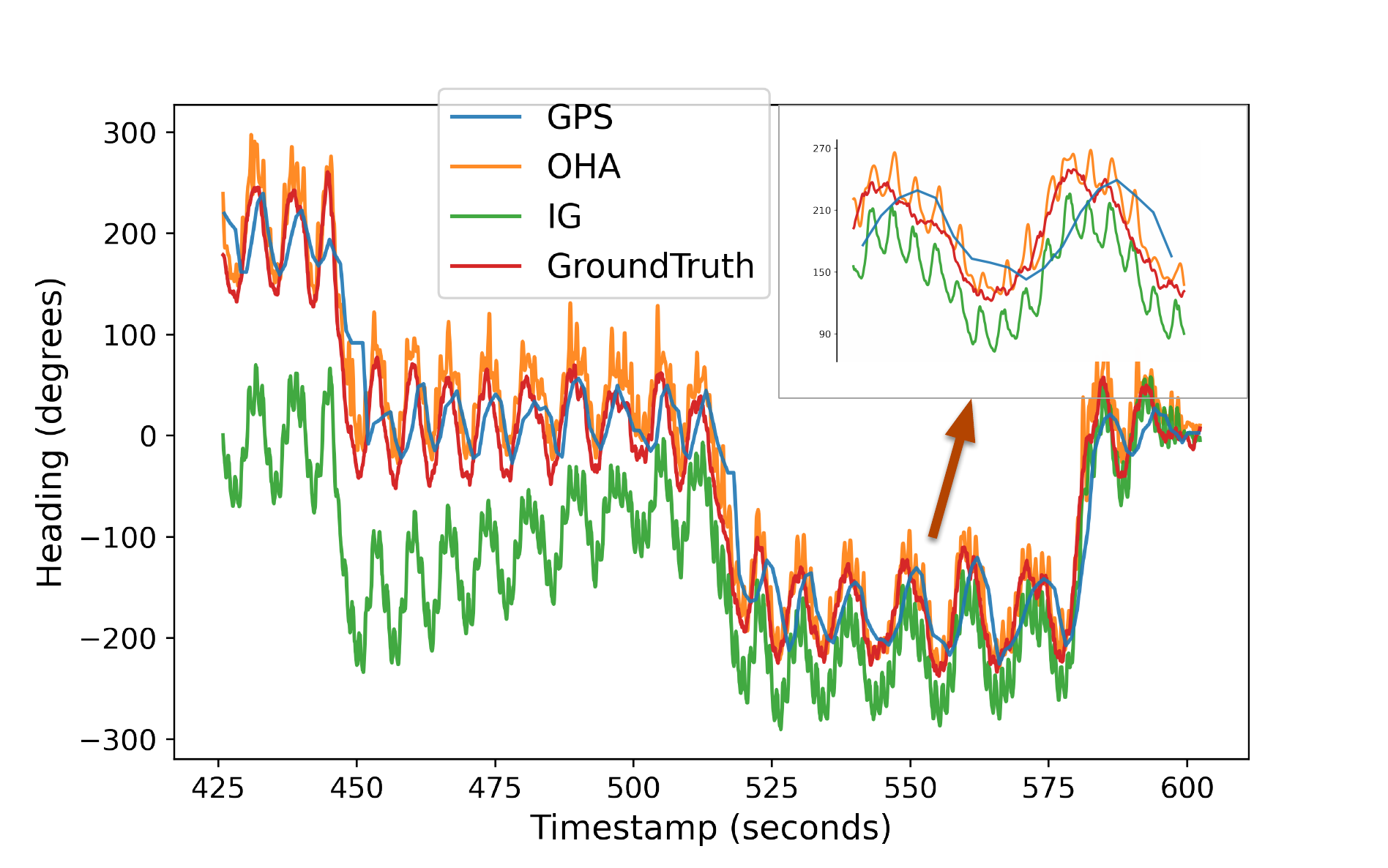}
        \caption{}
        \label{fig:heading_demo}
        \Description{}
    \end{subfigure}
    \hfill
    % Second figure: Magnetic Field Distribution
    \begin{subfigure}[b]{0.18\linewidth}
        \includegraphics[width=\linewidth]{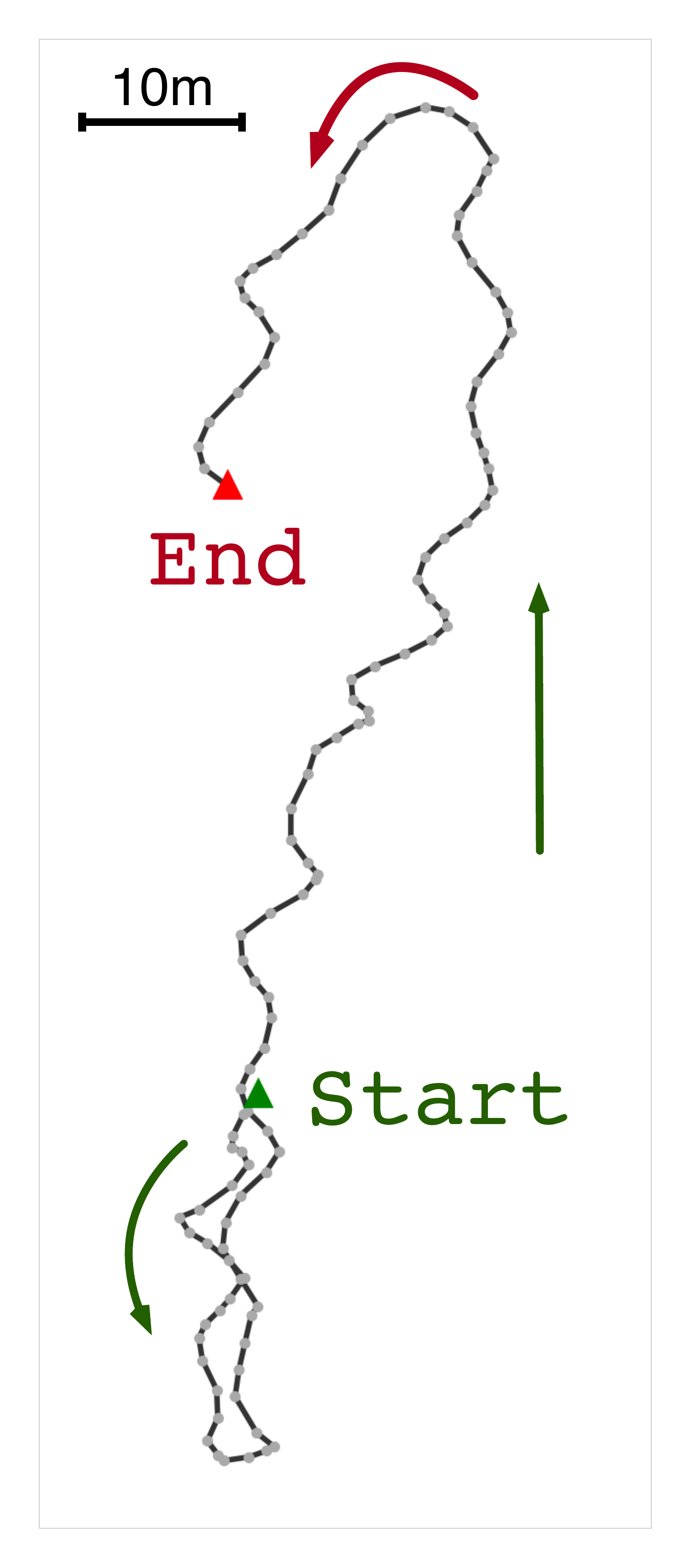}
        \caption{}
        \label{fig:heading_demo_trace}
    \end{subfigure}
    \caption{Participant places the smartphone in his trouser pocket and walks in ``S''-shaped paths. (a) A sample trace of heading over time for three heading estimation techniques compared with ground truth heading. (b) walking trajectory of the sample trace in (a).}
\end{figure}

% \begin{figure}[ht]
%     \centering
%     % First minipage for figures A and B
%     \begin{minipage}{0.55\textwidth}
%         \centering
%         \begin{subfigure}[b]{0.78\textwidth}
%             \includegraphics[width=\textwidth]{Figures/demo_sample.pdf}
%             \caption{}
%             \label{fig:heading_demo}
%             \Description{}
%         \end{subfigure}
%         \hfill
%         \begin{subfigure}[b]{0.18\textwidth}
%             \includegraphics[width=\textwidth]{Figures/sample_trajectory.pdf}
%             \caption{}
%             \label{fig:heading_demo_trace}
%             \Description{}
%         \end{subfigure}
%         \caption{Participant places the smartphone in his trouser pocket and walks in ``S''-shaped paths. (a) A sample trace of heading over time for three heading estimation techniques compared with ground truth heading. (b) walking trajectory of the sample trace in (a).}
%     \end{minipage}%
%     % Second minipage for figure C
%     \hfill
%     \begin{minipage}{0.44\textwidth}
%         \centering
%         \includegraphics[width=\textwidth]{Figures/gps_accuracy_error_bar.png}
%         \caption{OHA heading errors related to different levels of GPS signal blockage. The smartphone was placed in a box lined with aluminum foil, with the opening controlled as follows: (1) Normal - no cover, (2) Open - with an opening, (3) Half-Open - with the opening partially closed, (4) Closed - completely closed.}
%         \label{fig:gpsacc}
%     \end{minipage}
% \end{figure}

Figure \ref{fig:heading_demo} compares three heading estimation techniques over a 3-minute sample trace. Before this sample trace, the participant has walked for 7 minutes with the smartphone in his trouser pocket, and continues walking in ``S''-shaped paths as illustrated in Figure \ref{fig:heading_demo_trace}. During certain short intervals (e.g., 30 seconds), the IG method can effectively mirror the heading change pattern and achieve minimal errors as in previous works~\cite{Manos2018gravity, thio2021relative}. However, it suffers from uncontrollable cumulative errors over longer periods, which causes deviations from the ground truth headings. Meanwhile, the GPS heading, while free from drift errors, exhibits a consistent response delay of approximately 2 seconds when following the participant's turns. In the contrast, OHA demonstrates its ability to accurately track the pedestrian heading over a long time even when the user turns frequently.  

\begin{figure}[ht]
    \centering
    \includegraphics[width=0.43\textwidth]{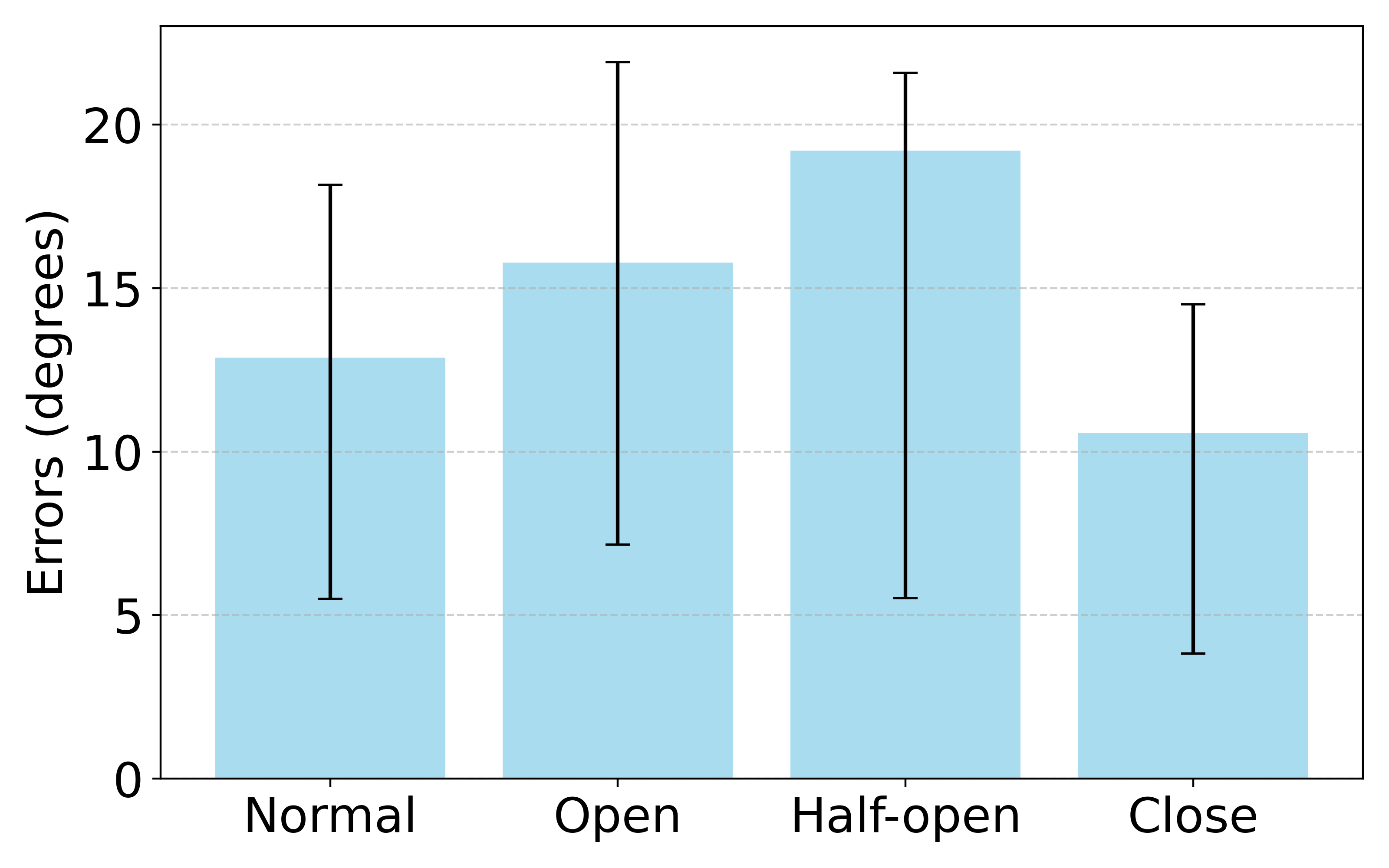}
    \caption{OHA heading errors related to different levels of GPS signal blockage. The smartphone was placed in a box lined with aluminum foil, with the opening controlled as follows: (1) Normal - no cover, GPS accuracy is around 4 meters, (2) Open - with an opening, (3) Half-Open - with the opening partially closed, (4) Closed - completely closed.}
    \label{fig:gpsacc}
\end{figure}

\noindent\textbf{GPS Accuracy:} We further evaluate the impact of GPS accuracy on OHA’s performance. To simulate GPS signal interference, we placed the smartphone inside a box lined with aluminum foil, and varied the degree of signal blockage by adjusting the box’s gap, from fully open to completely closed. Throughout the experiment, the smartphone was kept in the box at all times, and we carried the box while walking in two patterns: Straight and Occasional Turns (SOT) and Multiple S-shaped Paths (MSP). We observed that with the box fully closed, the GPS accuracy was consistently around 7.3 meters while occasionally reaching up to 20 meters. Figure \ref{fig:gpsacc} illustrates a comparison of OHA’s performance across different GPS accuracy conditions. Despite variations in GPS accuracy, OHA maintained satisfactory performance, demonstrating its potential for real-world environments. 

\section{Part II: Pedestrian Road Crossing Prediction} \label{sec:crs_pred}
Precise pedestrian heading is crucial for various applications, including enhancing crosswalk safety in urban environments, offering detailed navigation assistance in both indoor and outdoor environment, and improving the accuracy of location-based services on smartphones. 

To demonstrate our technique in a practical scenario, we apply OHA  in one of the more challenging tasks: predicting pedestrian’s crossing behavior with \textbf{low latency} and \textbf{high accuracy} using commodity devices. GPS bearing is inadequate for this task due to its inaccuracy and inherent delays. Additionally, it fails to determine pedestrian’s heading when they are stationary and rotating, a common scenario when a pedestrian decide to cross. We introduce \name, a system that predicts pedestrian’s crossing behavior based on pedestrian’s precise heading from OHA and pedestrian’s geographical location on roads.

\subsection{\name Overview}

\begin{figure}
    \centering
    \includegraphics[width=0.8\linewidth]{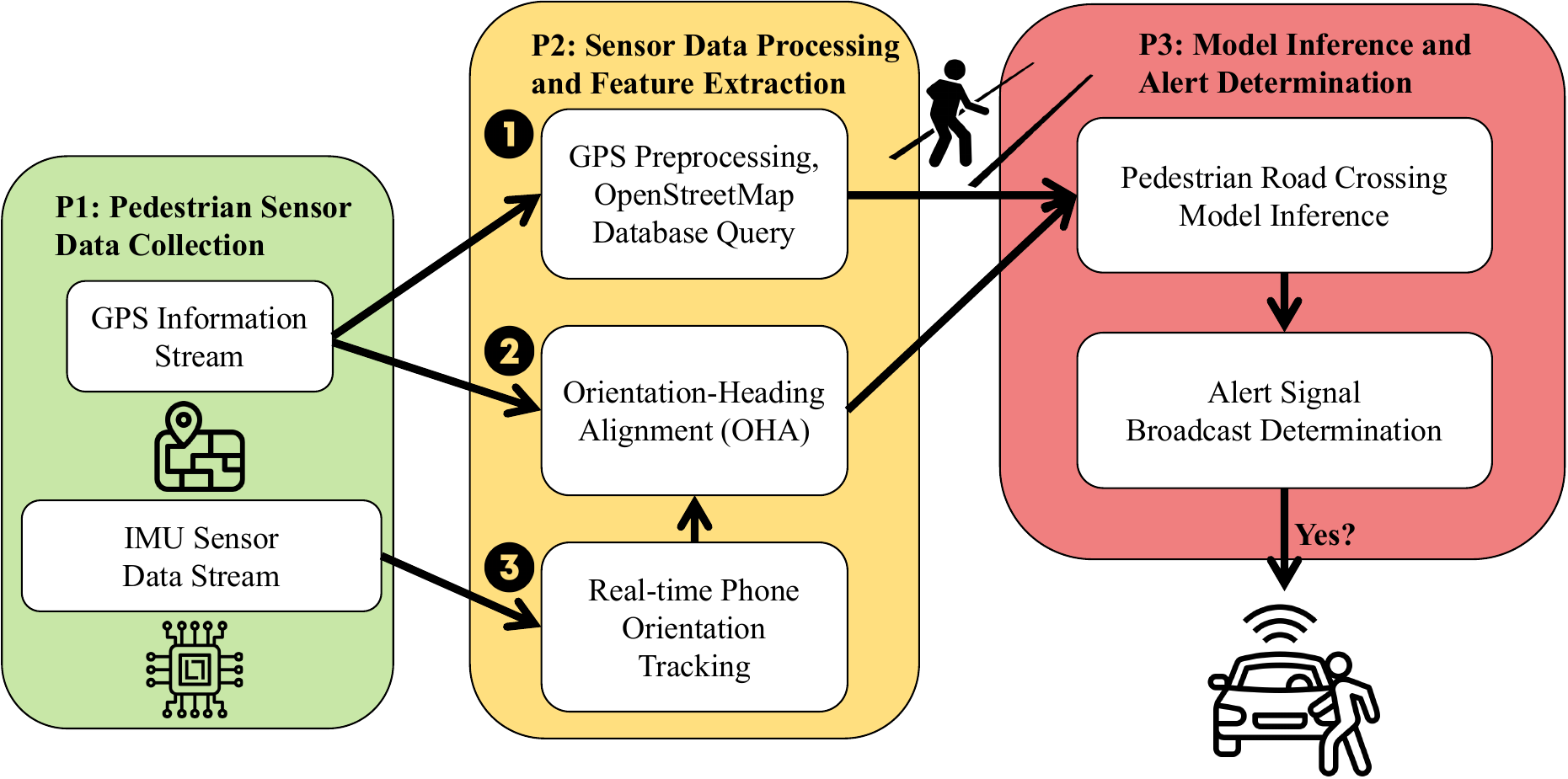}
    \caption{Pedestrian Road Crossing Prediction system overview. Phase 2 accepts GPS and IMU sensor data from Phase 1, and delivers Phase 3  1) general physical location and 2) OHA heading. We will introduce the road crossing model development and alert signal determination in Section \ref{sec:feature_extract}, and implementation details in Section \ref{sec:impl}.}
    \label{fig:sys_diagram}
    \Description{}
\end{figure}

\name aims to predict instances when a pedestrian is \textit{about to cross the road closest to them} and broadcasts alert messages to warn nearby drivers. These instances include pedestrians crossing at traffic lights or jaywalking at road mid-blocks. Figure \ref{fig:sys_diagram} provides a high-level overview of \name, illustrating its main components. At a high level, \name operates in three phases. In the first phase, it samples GPS and IMU data, which it uses in phase 2 to extract fine-grained mobility features about the user. In the third phase, it feeds these features to a model that predicts when the user is about to cross the road.

\noindent
\textbf{Phase 1:} \name utilizes GPS data and map information to identify the closest road to the user and estimate the coarse-grained distance to the road. This design makes \name resilient to on-device GPS inaccuracies, especially in urban environments~\cite{gpsaccuracy}. It is challenging to use GPS to detect real-time user movements, such as turning towards the road or stepping towards the road edge, with a submeter accuracy.

\noindent
\textbf{Phase 2:}
\name estimates two mobility features: coarse-grained distance to the closest road and the real-time heading from OHA. As evident in Section~\ref{sec:oha_accu}, \name overcomes the challenges of estimating the fine-grained heading information from IMU. Meanwhile, \name requests the nearby road shape data, to assist in identifying the intent of pedestrian road crossing.
% Note that phone measurements do not directly indicate a pedestrian's heading as they occur in the LCS and not in the GCS. Second, IMU sensor data is prone to drifts and noise~\cite{android:motionsensor}, which limits its usefulness for extended periods. \smallskip

\noindent
\textbf{Phase 3:} 
Finally,  \name feeds the recent history of these mobility features to a model, which predicts whether the user is about to cross the road. Incorporating recent mobility information gives the model more context for the road crossing prediction task. Moreover, the model can uncover some pedestrian mobility patterns that rule-based approaches can not~\cite {rudenko2020human}. Section \ref{sec:eval_q2} provides more insights into how the model uses the features.

We integrate the above functionality into a standalone and lightweight user-level app on the Android platform. This app requires no special hardware, infrastructure, or user support. It autonomously predicts when the user is about to cross the road and broadcasts a wireless message to alert nearby road users about potential road crossings. We base our broadcasting mechanism on the existing Wi-Fi Aware protocol through beacon stuffing~\cite{beacon_stuffing_paper}. Section \ref{sec:impl} contains more details about our implementation. This design makes \name easier to deploy and adopt than other solutions requiring dedicated infrastructure or special hardware.

\subsection{Machine Learning Model Details}\label{sec:feature_extract}
\name utilizes a machine learning model to predict whether a pedestrian will cross the road using recent mobility information as features. There are two challenges in predicting a pedestrian's road crossing: the lack of road width information and GPS inaccuracies. Both challenges prevent \name from determining the pedestrian location relative to the road. Even the fine-grained heading information from the OHA algorithm is not enough to decide on road crossing without some notion of the user's location. \name leverages an intuitive observation about pedestrian road-crossing behavior to tackle these challenges: \textit{pedestrians turn towards the road before crossing it}. By consolidating inputs from a pedestrian's recent OHA heading, location information, and map data, \name can assess the pedestrian's crossing potential.

%%%%%%%%%%%%%%%%%%%%%%%%%%%%%%%%%%%%%%%%%%%%%%%%
\noindent\textbf{Feature Extraction:} \name uses three key features: the distance to the road center, OHA heading (pedestrian heading from OHA), and the road reference angle. The distance information represents how close the pedestrian is to the nearest road. The OHA heading, along with the road reference angle, helps determine the user's movements relative to the closest road. We describe each feature used as follows:
\noindent
\begin{itemize}[leftmargin=*]

    \item \textbf{Distance to Road Center $d$} represents the perpendicular distance to the centerline of the nearest road, given the user's current GPS location. Despite imprecise GPS data and the lack of accurate road width information, the distance to the road center indicates whether a pedestrian is within a possible road crossing region. 

    \item  \textbf{Road Reference Angle $\theta_{ref}$} provides a reference to assess if a pedestrian is approaching or departing from a road. It represents the pedestrian's heading angle when they face perpendicular to the road. \name extracts $\theta_{ref}$ using the road shape in map data.
    
    \item  \textbf{OHA Heading $\theta_{oha}$}, the pedestrian's current heading in real-time, helps reveal the pedestrian's crossing potential along with $\theta_{ref}$. In particular, we use $\cos{(\theta_{ref} - \theta_{oha})}$, the cosine value between the OHA heading and road reference angle as a normalized input feature. 
    
    \item  \textbf{Prediction Window} represents the length of the input feature vector. In particular, the model input is a vector of the last \textit{N} readings of $\theta_{ref}$, $\theta_{oha}$, and $d$. Depending on the sampling frequency in the smartphone, this vector will represent the past \textit{lookback} seconds of mobility data. 

\end{itemize}

\noindent\textbf{Model Architecture:} We employ a dual LSTM (Long Short-Term Memory) layer configuration to efficiently process the extracted features on daily walking trace data. Specifically, one LSTM layer analyzes distance-related sequences and the other focuses on heading-related sequences, both operating in parallel to effectively capture temporal patterns. The outputs from these layers are subsequently concatenated and fed into a fully connected layer with a sigmoid activation function, producing the crossing probability at the given timestamp. 

\noindent\textbf{Crossing Alert Determination:} Rather than relying on the model's single prediction result, \name determines whether to broadcast a crossing alert after consulting the past $n$ predictions. This approach reduces the occurrence of false alarms that can undermine the user's confidence in \name. If the number of ``True'' crossing predictions in the past \textbf{$n$} predictions surpasses a threshold, usually 50\%, \name will broadcast a crossing alert to nearby vehicle drivers.
With a larger $n$, \name provides a more reliable, yet delayed, crossing alert. On the other hand, with a smaller $n$, \name issues an alert earlier, but with the risk of generating more false alarms. 
 
%%%%%%%%%%%%%%%%%%%%%%%%%%%%%%%%%%%%%%%%%%%%%%%%
\subsection{Data Workflow and Model Development}\label{sec:60p_collection}
We curated a dataset of walking traces from 60 individuals in their daily lives. Following a process of data cleansing, labeling, and feature extraction, we constructed and refined our model using this dataset.
% The model prediction is ``True'' if the probability is higher than 0.5. 

\subsubsection{Data Collection}
\label{sec:daily_wk_ds}
We collected traces of IMU and GPS data when participants were walking in their daily lives. Our data collection protocol was approved by our institute IRB. From 2020 to 2022, we recruited 60 participants from campus, who participated in the data collection from two weeks to two months. Our study began with a consent phase, where we gave an overview of the study and explained our data protection measures and participants' rights. 
Participants who consented to join the study installed our data collection app on their personal smartphones and kept it active in the background. We developed this app based on GPSLogger\footnote{https://github.com/mendhak/gpslogger}.

As our data collection might accidentally collect sensitive GPS information, we implemented multiple privacy protection measures. First, each participant can select points of interest (POI) to opt out of the collection, such as homes, offices, and other places. Second, our app provided an opt-out button to allow participants to manually stop data logging on demand. In addition, we removed any possible identifier from the uploaded data, associating the data only with hashed advertisement IDs.

We compensated each participant who completed the two-month session with a \$40 Amazon gift card. 
In total, we collected around 755 hours of walking data (2.72 million GPS data points and corresponding IMU data after filtering, which will be discussed in the following).

\subsubsection{Trace Segmentation and Labeling}
The daily walking traces comprise pedestrians' hours-long GPS and sensor data collected when they walked.
However, our GPS recording suffered from occasional GPS signal loss and was interrupted when participants opted out.
Therefore, for each pedestrian, we segmented all location traces by time intervals into smaller sessions, referred to as \textbf{road-use sessions}. 
Each road-use session represents a complete and continuous (with signal loss shorter than 15 seconds) outdoor walking event, which consists of GPS and sensory data. In total, we extracted around 2,000 road-use sessions, and each road-use session's length ranges from 1 minute to 28.6 minutes. 

The next task was labeling road crossings in the pedestrians' walking traces, which we found challenging for two reasons. First, it is difficult to determine every actual road-crossing behavior, with inherent GPS inaccuracies and unknown road width information.
Second, it is difficult to differentiate participants' potential road-crossing behaviors from GPS drifts. Therefore, we cleaned and labeled the traces to minimize the influences of GPS inaccuracies and drifts that bias the model, following the following steps:

    \textbf{Step 1.} We defined and labeled a ``true crossing event'' as the period from when a pedestrian \textbf{shows potential to cross} to when they \textbf{reach the road center}. Note that we do not include the time period after reaching the road center, as doing so may delay the early detection of pedestrian road crossing. We determined the start of crossing potential by detecting turnings from GPS bearing change if the pedestrian turned to cross. For those who did not turn, we defined it as occurring five seconds before the pedestrian reached the road center. We consider five seconds to be a reasonable timeframe for a pedestrian to reach the road center once they begin crossing.

\begin{figure}[ht]
    \centering
    % First figure: Participant Trace
    \begin{minipage}[b]{0.48\linewidth}
        \centering
        \includegraphics[width=\linewidth]{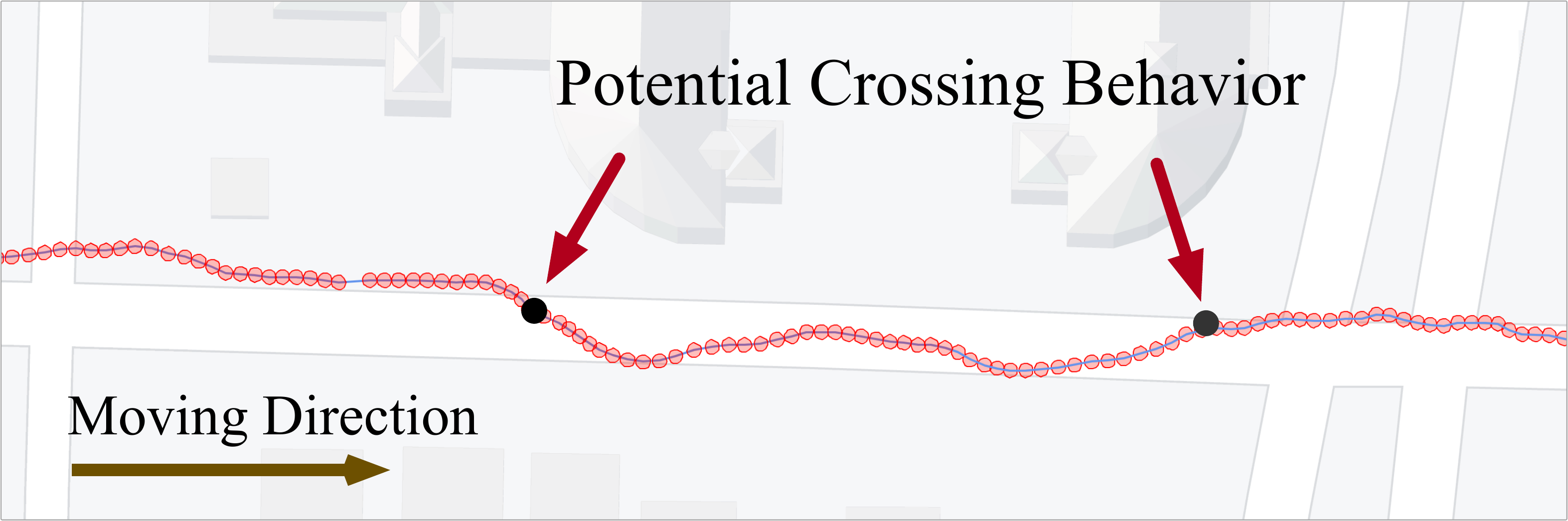}
        \caption{Participant walking trace and potential crossing behavior caused by GPS drift. ``Crossed'' twice in a short period.}
        \label{fig:participant_trace}
    \end{minipage}
    \hfill
    % Second figure: Magnetic Field Distribution
    \begin{minipage}[b]{0.48\linewidth}
        \includegraphics[width=\linewidth]{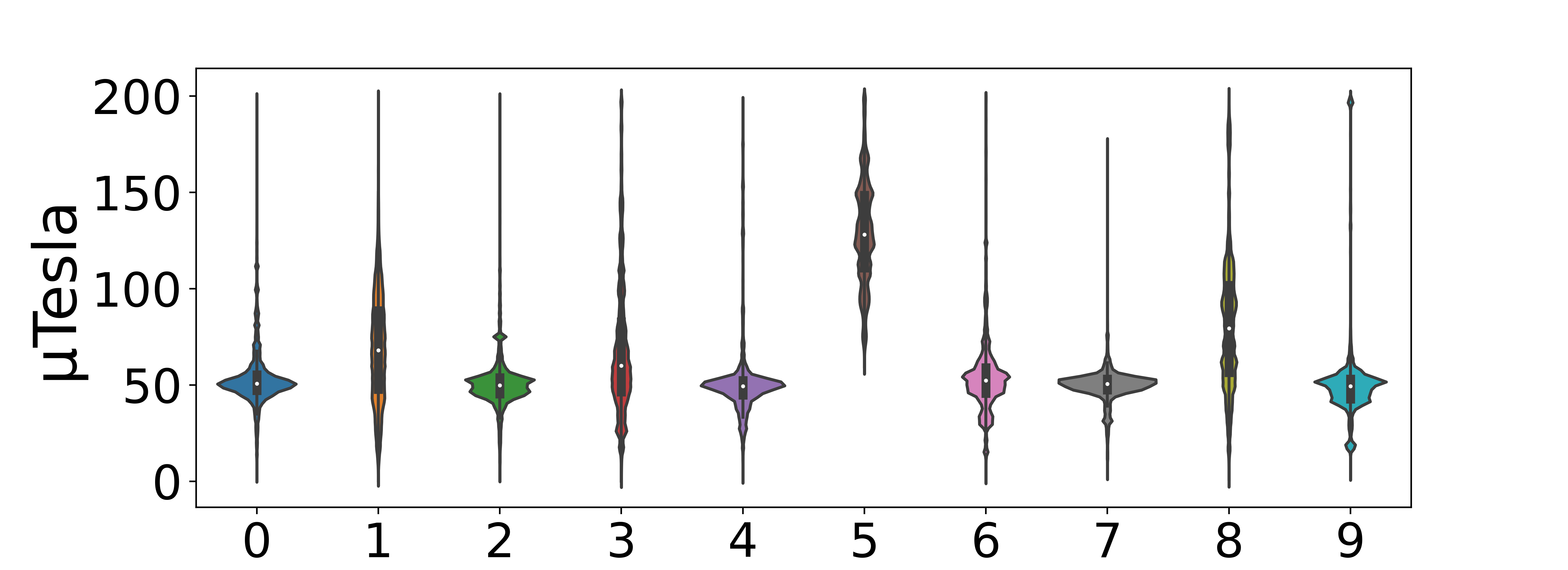}
        \caption{Distribution of Magnetic field magnitude measured by ten pedestrians while walking outdoors.}
        \label{fig:mag_dist}
    \end{minipage}
\end{figure}

    \textbf{Step 2.} We removed certain events where the road-crossing behavior is ambiguous, such as the example shown in Figure \ref{fig:participant_trace}. 
    Specifically, we ruled out the sessions where pedestrians returned to the original side of the road immediately after crossing.
    Note that our cleaning aims to label data with higher confidence. While map matching techniques~\cite{project-osrm} label two crossing events for the trace in Figure \ref{fig:participant_trace}, our cleaning method will remove such trace due to low confidence.

Our processed data comprises timestamped and labeled walking traces, including when the pedestrian is not crossing and when their crossing has high confidence.

% \begin{figure}
%     \centering
%     \includegraphics[width = 0.5\linewidth]{Figures/suspicious_crossing.pdf}
%     \caption{Participant's road-use-session trace and potential crossing behavior caused by GPS drift. The blue marker represents the trace start point. The participant ``crossed'' the road twice during a short time period.}
%     \label{fig:participant_trace}
% \end{figure}
% \begin{figure}[h]
%     \centering
%     \includegraphics[width =0.5\linewidth]{Figures/mag_norm_violin_10.png}
%     \caption{Distribution of Magnetic field magnitude measured by ten pedestrians while walking outdoors.}
%     \label{fig:mag_dist}
% \end{figure}

\subsubsection{Model Training and Testing Results}

We selected $lookback$ length as 8 seconds, extracting the input features and predicting every 100~ms ($N=80$). We use two 64-unit LSTM layers for processing the two feature vectors and a dense layer to process the concatenated 128-long LSTM outputs. For issuing a crossing alert, we choose $n=20$, equivalent to 2 seconds, to balance the trade-off between crossing alert timing and false alarm rate. 

We implemented our model using Tensorflow and trained it on four NVIDIA RTX 2080 Ti GPUs. During training, since temporally close samples exhibit strong similarity, we do not shuffle the samples in the training set. We set aside 10\% of the training data as the validation set. We also use early stopping with patience as 3 to avoid model overfitting. Note that the samples in our daily walking traces are imbalanced, consisting of roughly 5\% true crossing events and 95\% non-crossing traces. As such, we employ precision and recall on true crossing events as our evaluation metrics.

\begin{table}[t]
\centering
\begin{tabular}{lccccc}
\toprule
\textit{lookback} & t = 2s & t = 4s & t = 8s & t = 16s \\ \midrule
Precision (\%) & 85.40 & 83.97 & \textbf{85.61} & 82.62 \\
Recall (\%) & 90.85 & 89.79 & \textbf{89.59} & 87.65 \\
\bottomrule
\end{tabular}
\caption{Precision and Recall rate on the test dataset of models trained with varying \textit{lookback} lengths.}
\label{tab:vary_length}
\end{table}

\noindent \textbf{Road Crossing Prediction Results:} We tested the road crossing prediction model on the granularity of events instead of each input sample. Originally, the model predicts the road-crossing probability for each sample in the road-use session, resulting in a probability sequence. We determine crossing alerts on this sequence and retrieve crossing alert periods with start- and end-timestamps. We test these periods against the labeled ``True'' crossing events in the road-use session. We define a true alarm as an overlap between a crossing alert period and a ``True'' crossing event, while the absence of such an overlap is a false alarm.

For additional testing, we trained four models with different \textit{lookback} lengths, including 2s, 4s, 8s, and 16s. We present each model's performance on the testing dataset in Table \ref{tab:vary_length}.
The recall rate keeps decreasing as \textit{lookback} length increases, while precision and recall rate drop to the lowest when the model utilizes past 16s sensor information.
\name choose \textit{lookback} length as 8s which yields highest precision rate (85.61\%) among four models. 
We also compared with baseline models in Section~\ref{sec:baseline}.

\subsection{Implementation Details}\label{sec:impl}
We implement \name as a standalone app on the Android platform. In the following, we briefly mention some implementation details and describe our crossing alert broadcasting mechanism.

% \begin{figure}[h]
%     \centering
%     \includegraphics[width =0.5\linewidth]{Figures/mag_norm_violin_10.png}
%     \caption{Distribution of Magnetic field magnitude measured by ten pedestrians while walking outdoors.}
%     \label{fig:mag_dist}
% \end{figure}

\noindent \textbf{Orientation Estimation Modifications: }
Figure~\ref{fig:mag_dist} shows the outdoor magnetic field distributions for ten users from the daily walking traces described in Section~\ref{sec:daily_wk_ds}. We can observe significant variations in magnetometer readings both across devices and over time, which affects the performance of MUSE. To mitigate this, we empirically constrained the range of magnetometer sensor readings, which effectively reduced the maximum orientation error caused by environmental magnetic fields.

%% remove wireless part for IMWUT2024
% \noindent\textbf{Broadcasting Alerts:}
% We build the crossing alert broadcasting mechanism in \name on Wi-Fi Aware protocol~\cite{wifiaware}, also called Neighbor Awareness Networking (NAN). This protocol allows for peer-to-peer connections without the need for a traditional Wi-Fi network or access point. In Wi-Fi Aware, there are two key roles, publisher and subscriber, which correspond to pedestrians and drivers, respectively, in our context. A pedestrian, acting as the publisher, transmits Service Discovery Frames (SDFs) containing service information. This enables drivers, as subscribers, to detect the presence of pedestrians. Instead of establishing connections, pedestrian embeds crossing alert (up to 255 bytes) inside this SDF, also called beacon stuffing~\cite{beacon_stuffing_paper}. Therefore, a pedestrian who is going to cross can publish a crossing alert to specific nearby drivers, who are actively listening for a matching publisher. 

% To minimize the discovery latency, we assign specific service identifiers and matching filters for pedestrians and drivers. Therefore, nearby users without these specific service names or matching filters will not be notified, avoiding spamming all users in the proximity. 

\noindent \textbf{\name Running Modes:}
For power-saving purposes, \name has two running modes: one is active mode, and the other is idle mode. In the active mode, \name operates IMU sensor data fetching and feature data processing at 50 Hz, GPS updates at 1 Hz, and model inference every 100 ms. In the idle mode, \name only retrieves GPS data and queries the OSM database at a relatively low frequency, such as every 2 seconds. In our implementation, \name enters the idle mode when the pedestrian is 10 meters away from the road or when the user is immobile for more than 1 minute, where the pedestrian indicate a low probability of crossing.

\noindent \textbf{App Integration:} We implemented \name as an Android app compatible with Android 11 to Android 13 devices. The app uses OpenStreetMap (OSM~\cite{osm_data}) for app data containing road shape information. We load the OSM data on a local Postgres database on the phone using Termux~\cite{app:Termux}. The app queries the database for the nearest road, the distance to the road center, and the road's angle using PostGIS~\cite{postgis}. It utilizes TensorFlow Lite for model inference at run-time. Finally, it uses Android's WifiAwareManager~\cite{android-wifi-aware-2023} to broadcast the road crossing alerts.

\section{Evaluation of Pedestrian Crossing Prediction} \label{sec:eval}
Evaluating pedestrian road crossing prediction includes two key aspects. First, we assess how accurately \name can \textbf{identify the crossing events} for new pedestrians. Specifically, we examine how \name-issued crossing alerts match the true crossing events. Second, we consider the promptness of these alerts, specifically how early \name can alert before a pedestrian crosses the road edge, referred to as ``\textbf{time-to-crossing}'' in previous research~\cite{fang2019intention, palffy2019occlusion}. We address these two key aspects respectively in Sec.~\ref{sec:eval_q1} and in Sec.~\ref{sec:eval_q2}. 
We designed the evaluation to answer the following questions, and we summarize our observations:

\begin{enumerate}[label={\textbf{Q\arabic*.}}]

    \item \textbf{[Crossing Identification] How well can \name generalize to new participants with fine-grained labeling?}
    We conducted a study that monitored another 25 participants for evaluating \name in a semi-controlled real-world environment. We took videos of walking participants on a designated route and the road environment, and then we recruited annotators to label these videos independently. Our labels include both pedestrian's completed crossing and crossing potentials. Overall, \name achieved a precision of 86.9\% and a recall of 93.6\% in identifying the completed crossings.

    \item \textbf{[Detailed Analysis] How early can \name issue alerts, what contributes to \name's error and how is \name compared to a baseline model?} 
    Comparing with fine-grained labels from our annotators, we observed that \name typically issues crossing alerts before a pedestrian crosses the road edge. We identified five error sources of \name, among which GPS errors and ambiguous pedestrian crossing intent contribute the most. We also illustrated six walking traces with marked predictions in Figure \ref{fig:prediction_sample}. After training the baseline models following the same procedures, we found that the baseline model issues a crossing alert 2.1 seconds after a pedestrian crosses the road edge.
    
    \item \textbf{[System overhead] What is \name's system overhead?} We measured the performance impacts of our system on multiple commodity devices with different Android versions. We found the average execution time for each running loop is around 10 ms, and the average battery consumption over 30 minutes is around 352 mAh for active mode and 95 mAh for idle mode. 
    % We also observed that the majority of alert broadcasts experience a delay of less than 400 ms and achieve a success rate of over 90\%.
    % IMWUT2024

\end{enumerate}

\subsection{\textbf{Q1. Crossing Identification of \name}}\label{sec:eval_q1}
To understand how well \name generalizes to new participants, we first conducted a user study with another 25 participants, separate from the 60 participants discussed in Sec. \ref{sec:60p_collection}. Then we evaluated how accurately can \name identify crossing events using data collected from this user study. 

\subsubsection{Semi-controlled Walking Trace Collection}
We conducted this study in a semi-controlled real-world environment and video-recorded the participants' walking behavior for labeling.
We designated the start and end points on our campus for participants to walk. When a participant was walking, we recorded the pedestrian and the road environment. This data collection received an exemption from our institute's IRB. The data collection took place from May to July 2023, and we recruited 25 participants (our institute's staff and students) across campus. Of the participants, 17 identified themselves as male (8 female). Participants' ages ranged from 20 to 45 years old.

We asked each of them to follow the same walking path in the campus area, a walkable environment, while carrying a phone with \name installed. Each participant walked at their own speed and decided to cross the road when they felt comfortable. The walking took around 15 minutes for each participant. A researcher followed and recorded participants' walking behaviors without interfering with them. We compensated each participant with office supplies worth \$8.
% \todo{Fixed. Add privacy and safety measures? why irb exemption?}

\begin{figure}
     \centering
     \begin{subfigure}[b]{0.3\linewidth}
         \centering
         \includegraphics[width=\linewidth]{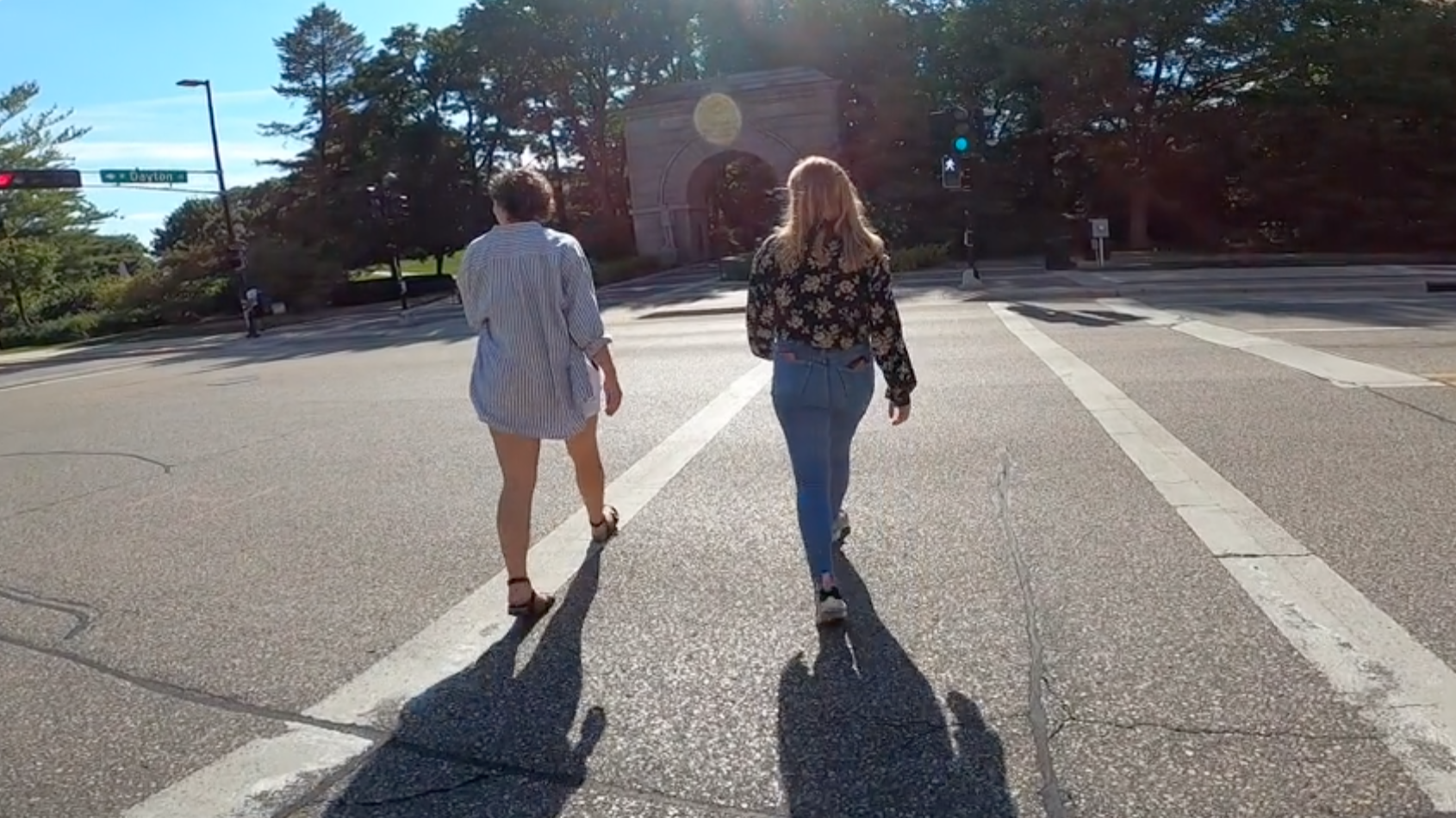}
         \caption{ac: actual crossing}
         \label{fig:ac}
     \end{subfigure}
     \begin{subfigure}[b]{0.3\linewidth}
         \centering
         \includegraphics[width=\linewidth]{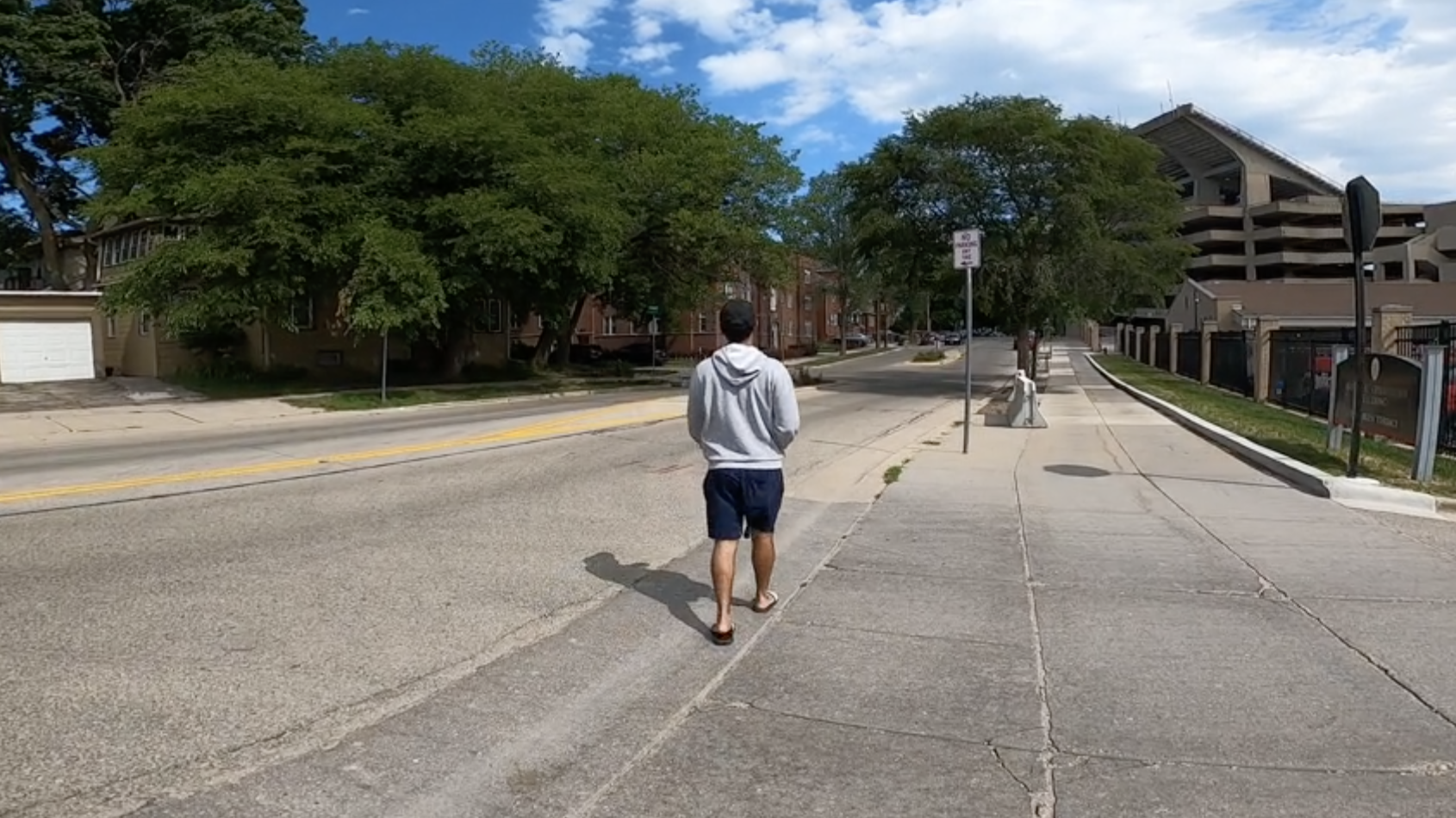}
         \caption{pc: potential crossing}
         \label{fig:pc}
     \end{subfigure}
     \begin{subfigure}[b]{0.3\linewidth}
         \centering
         \includegraphics[width=\linewidth]{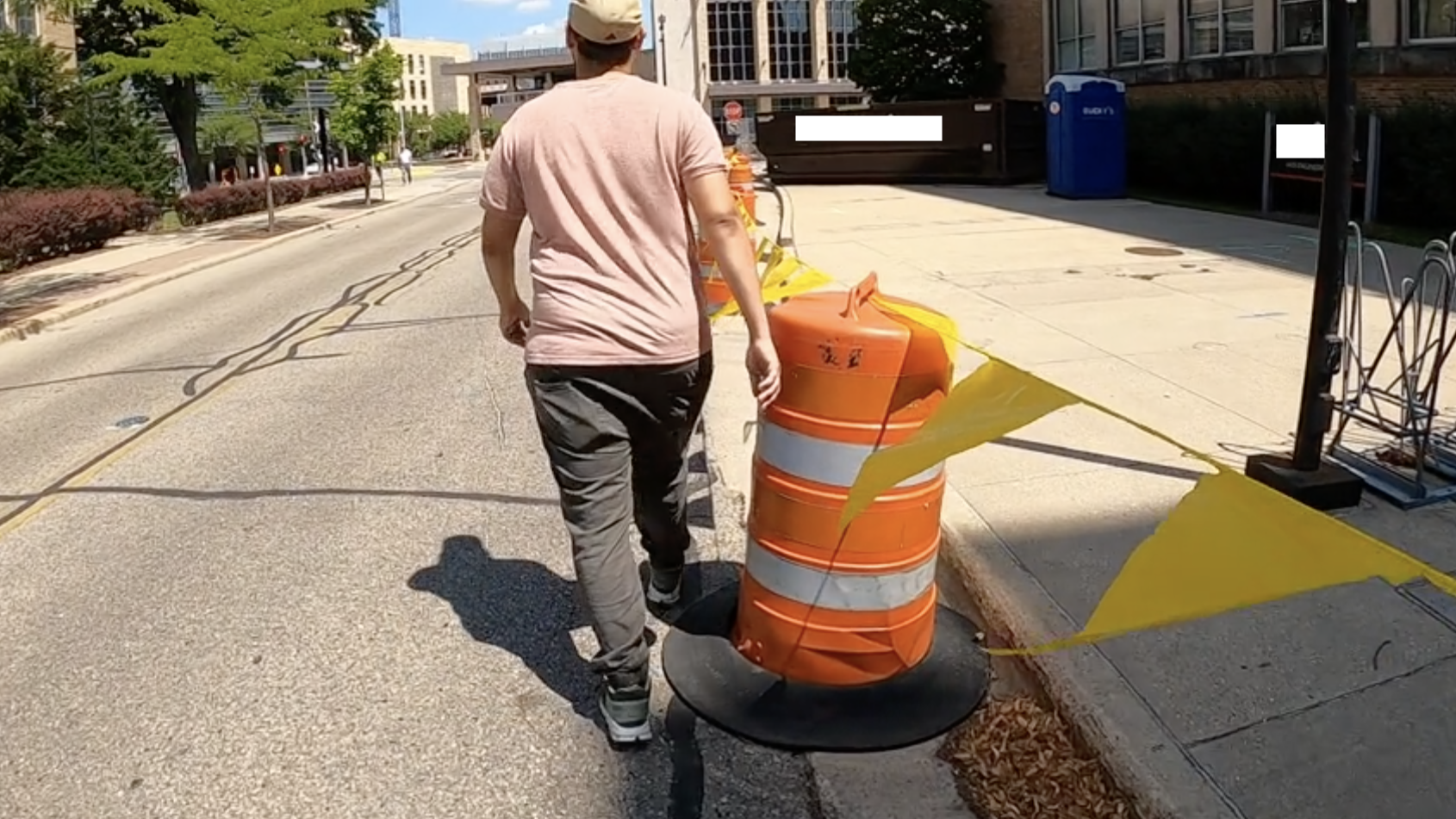}
         \caption{db: dangerous behavior}
         \label{fig:db}
     \end{subfigure}
        \caption{Examples of three behaviors, (a) actual crossing (ac), (b) potential crossing (pc), and (c) dangerous behavior (db) labeled by both annotators.}
        \label{fig:crossing}
        \Description{}
\end{figure}

\noindent \textbf{Video-based Road Crossing Labeling:}
We recruited two graduate students majoring in traffic safety to label and timestamp participants' walking behavior by watching the recorded videos. We identified three types of pedestrian behavior that need to be labeled: actual road crossing, potential road crossing, and dangerous/unsure behavior. Both annotators followed the labeling guidelines below:

    $\bullet$ \textbf{Completed/Actual road crossing:} Label this behavior when observing that a pedestrian has crossed the road. Mark the timestamps when the pedestrian (1) demonstrates crossing potential, (2) reaches the road edge and prepared to cross, and (3) departs from the road edge. 
    
    $\bullet$ \textbf{Potential road crossing:} Label this behavior when observing a pedestrian showing intent to cross but not crossing eventually. Mark the timestamps when the pedestrian (1) shows and (2) no longer shows crossing intent.
    
    $\bullet$ \textbf{Dangerous/Unsure behavior:} Label this behavior when observing a pedestrian displaying dangerous behavior or when being unsure if the participant demonstrates ``potential road crossing.'' Mark the timestamps when such behavior begins and ends.

Before labeling the whole walking traces, we asked both annotators to independently label a recorded walking video of one researcher to measure the agreement between them. 
We found that both annotators showed consensus on the occurrences of completed road crossing events but had minor disagreements on the other two behaviors. Their labeling results reached a high agreement with an Intersection over Union (IoU) value of 0.847. We present examples of three behaviors annotated by both annotators in Figure \ref{fig:crossing}.

% \subsubsection{Performance Evaluations}
\subsubsection{Crossing Identification Evaluation Results}

We evaluated \name's performance in identifying completed road crossings on the \textbf{semi-controlled walking traces}. Compared to daily walking traces discussed in Sec. \ref{sec:60p_collection}, semi-controlled walking traces further recorded detailed crossing behaviors, including participants' body movements and postures, to clearly determine when pedestrians reach the road's edge and prepare to cross. Through video-recording and fine-grained labeling, we can better evaluate \name's performance in identifying road crossings and validate our model's generalizability to new participants.
% We also trained baseline models with alternative input features for ablation study. 
% \todo{be more positive, you dont want to attack your own work. say something like: though our daily dataset contributes large amount of data, such experiment cannot record...}

% One weakness of evaluation on pedestrian's road crossing behavior in daily life is that we determine the start of pedestrian's road crossing based on GPS information and Google Map. We lack the ground-truth timestamps of pedestrian entering road range and miss the detailed understanding of \name 's performance in terms of early warning time and environmental noise impacts.

% Therefore, in the user-based evaluations we follow and record the pedestrian and traffic for fine-grained labeling.

% \subsubsection{Crossing Identification Evaluation Results}
\noindent\textbf{Performance Results:} Based on the labeling results from both annotators, we observed 443 complete crossings in total. The annotators were in full agreement about the completed crossing events. Regarding crossing potentials, one annotator identified 63 instances while the other identified 43.

We evaluated \name's prediction results using the complete crossings as ground truth. Specifically, we used complete road crossings as ``True'' crossing events and took the average of the precision and recall rates across the 25 participants.
Overall, our model achieved an 86.7\% precision rate and a 93.6\% recall rate in identifying participants' complete road crossings. It is important to note that we labeled the semi-controlled walking traces via examining video recording, while the crossing events in the daily walking traces were labeled using GPS, which inevitably includes false crossing labels. Therefore, although our model has never encountered these semi-controlled walking traces before, it exhibits better performance in the evaluation, demonstrating its generalizability to new participants.

% Why did the model perfprm better here than the dataset???

% (We allowed disagreement in the labels as potential road-crossing behaviors can be ambiguous in their nature.)
% \todo{Fixed. check this}

 % \smallskip 
% \todo{Fixed. connect above. how does the potential road crossing and dangerous behavior connect to here???}

% In Section \ref{sec:err_analy}, we will discuss in detail about false alarms root causes during evaluations. 

\subsection{\textbf{Q2. Detailed Analysis for \name}}\label{sec:eval_q2}
We performed a detailed analysis for \name in terms of time-to-crossing (TTC), which we define as the time from \name's alert start timestamp to pedestrian's crossing start timestamp. Further, we analyzed how \name performs when encountering various factors, e.g., the potential crossings, in error analysis. Additionally, we compared \name's performance against baseline models.  

\subsubsection{Time-to-crossing}
Time-to-crossing indicates how early \name can alert vehicle drivers about pedestrian crossing nearby. 
In particular, TTC represents the time difference between when \name issues an alert and when the participant enters the road range (``reaching road edge'' timestamp tagged by annotators). 
An earlier crossing alert provides drivers with more time to avoid potential collisions.

For each correctly identified crossing event, we calculate the time-to-crossing as 
We compared \name's alert time with both annotators' results respectively. When consulting the last $n=20$ predictions, equivalent to the past 2 seconds, \name issues an alert, on average, 0.39 and 0.32 seconds \textbf{ahead of} the ``reaching road edge'' event. 
Note that we do not include any road width information during training, and the roads traversed consist of 2-lane and 4-lane roads with varying road widths.

We also evaluated the trade-off between time-to-crossing and false alarm rate in Table \ref{tab:eval_n}. With a higher $n$, \name considers a longer past time period and more prediction results, leading to more confident decisions on whether to issue a crossing alert but also delaying the alert time.  In summary, a higher $n$ increases the precision rate but decreases the time-to-crossing. Furthermore, in Table~\ref{tab:time_to_crossing}, \name demonstrates performance comparable to camera-based pedestrian detection techniques, while maintaining a relatively high precision rate and functioning effectively in NLOS situations.

\begin{table}
\centering
\begin{tabular}{lcccccc}
\toprule
Past $n$ predictions & 10 & \textbf{20} & 30 & 40 & 50 \\
\midrule
Precision (\%)		& 83.8	& \textbf{86.7}	& 89.6	& 91.6	& 91.3 \\
Time-to-crossing (s)	& 0.62 & \textbf{0.36} & -0.43 & -0.85 & -1.54 \\
\bottomrule

\end{tabular}
\caption{Trade-off between time-to-crossing and false alarm rate by varying past consulted predictions $n$. 10 predictions equal to 1s and negative TTC represents crossing alerts happen after pedestrian has crossed the road edge.}
\label{tab:eval_n}
\end{table}

% Meanwhile, we observed differences in annotators' understanding of ``reaching road edge'' timestamp. One reason is that the road boundaries were sometimes not clearly set and visible.
% For example, the two annotators disagreed on when to label the ``reaching road edge'' behavior when the participant walked across the road shoulder and stayed by. \smallskip
% \todo{Fixed. pls check this}
% one pedestrian crossed road shoulder and waited by the street parking stalls. 
% One annotator labeled the timestamp when the pedestrian crossed the road shoulder as ``reaching road edge'', while the other annotator chose the timestamp when the pedestrian started crossing after waiting.
% The first annotator indicates that ``when a pedestrian steps out of the sidewalk and puts a foot on the road, it signifies that they have reached the road edge, indicating their intention to cross''.

\subsubsection{Error Analysis}
\begin{figure}[ht]
    \centering
    \begin{subfigure}{0.3\textwidth}
        \includegraphics[width=\linewidth]{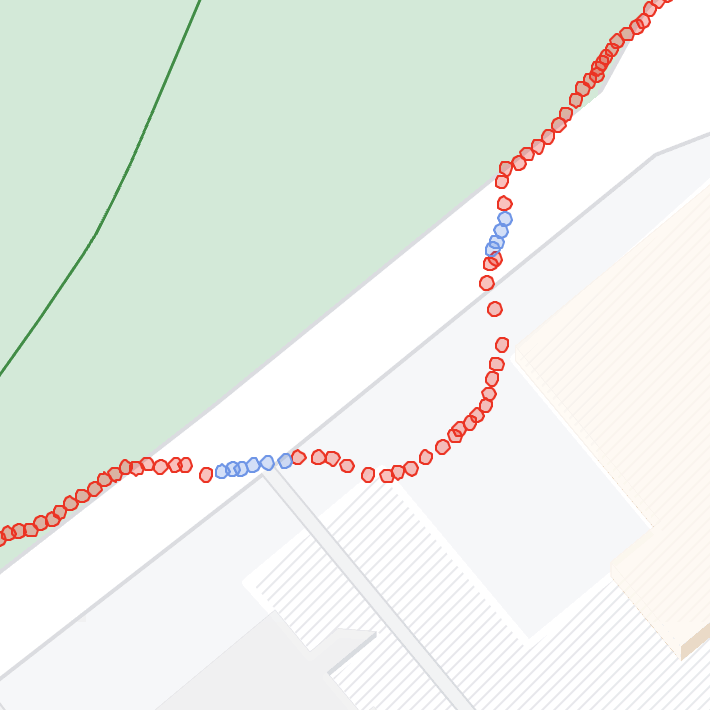}
        \caption{}
        \label{fig:subfig1}
    \end{subfigure}
    \hfill
    \begin{subfigure}{0.3\textwidth}
        \includegraphics[width=\linewidth]{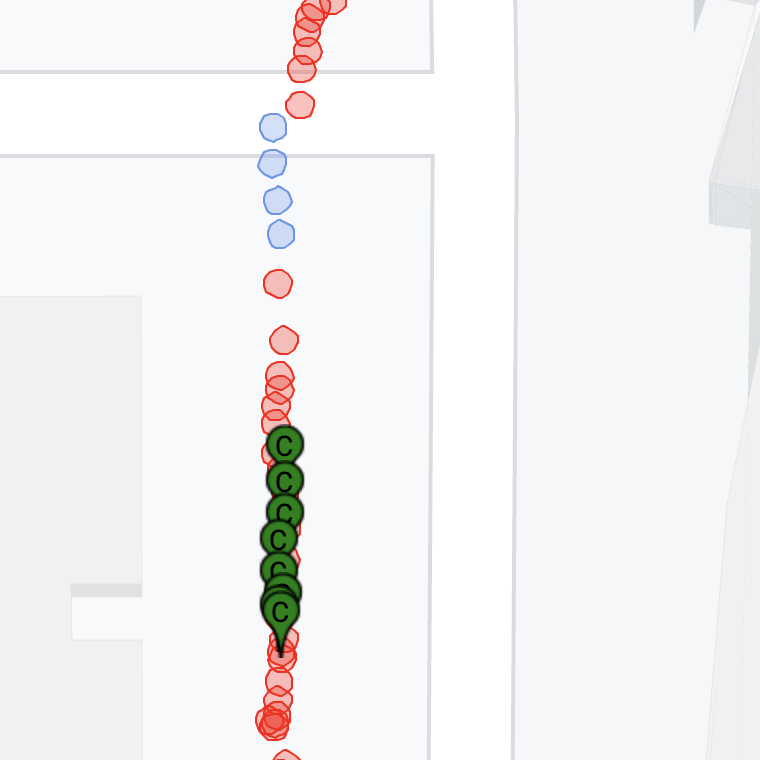}
        \caption{}
        \label{fig:subfig2}
    \end{subfigure}
    \hfill
    \begin{subfigure}{0.3\textwidth}
        \includegraphics[width=\linewidth]{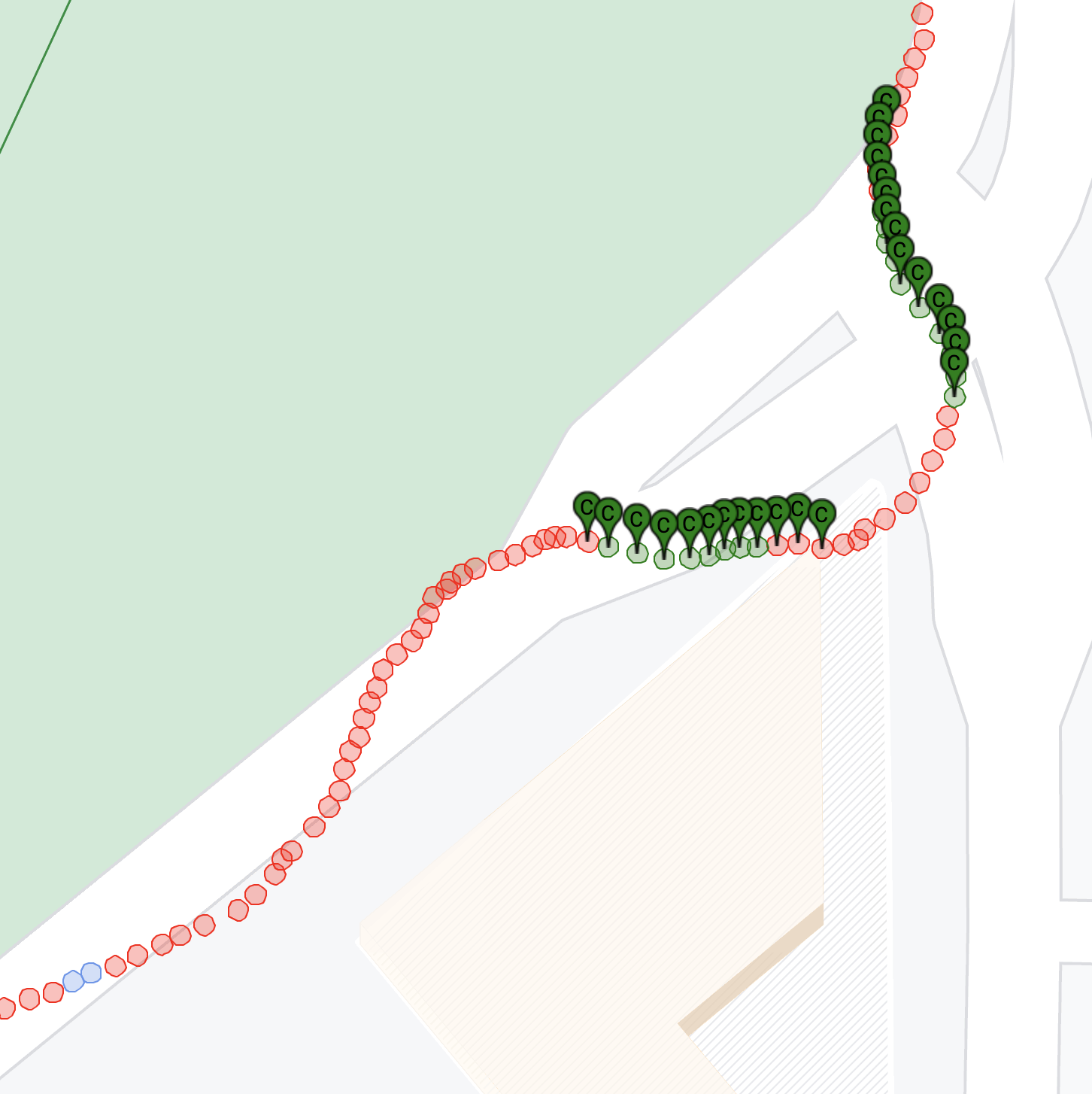}
        \caption{}
        \label{fig:subfig3}
    \end{subfigure}

    \begin{subfigure}{0.3\textwidth}
        \includegraphics[width=\linewidth]{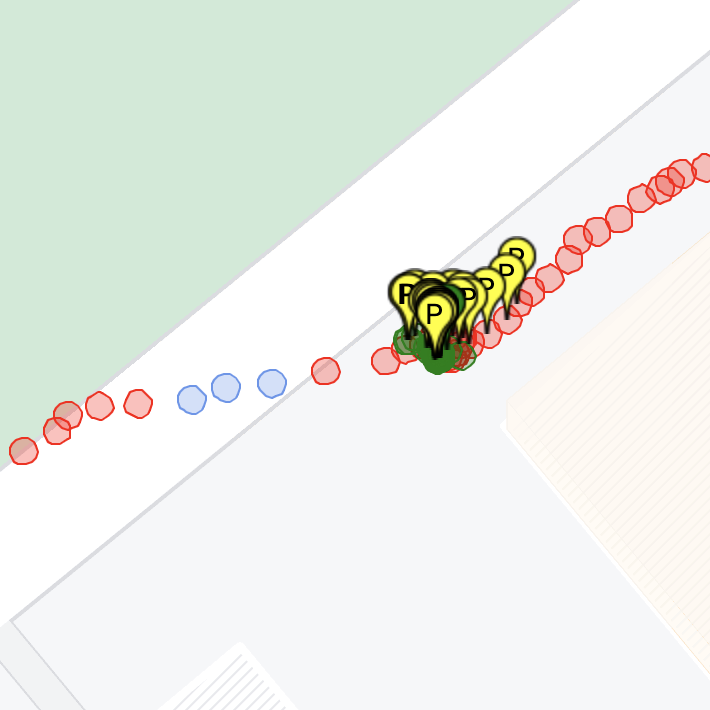}
        \caption{}
        \label{fig:subfig4}
    \end{subfigure}
    \hfill
    \begin{subfigure}{0.3\textwidth}
        \includegraphics[width=\linewidth]{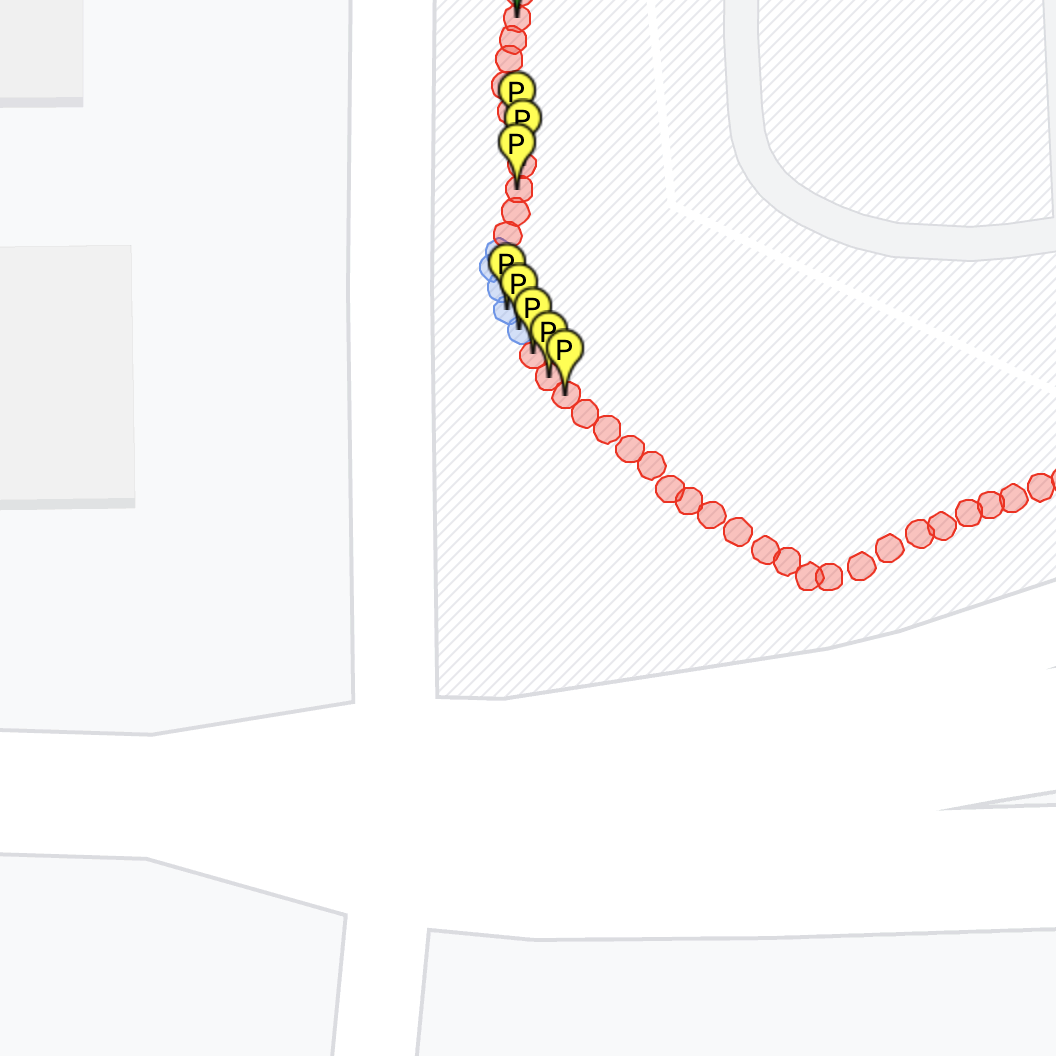}
        \caption{}
        \label{fig:subfig5}
    \end{subfigure}
    \hfill
    \begin{subfigure}{0.3\textwidth}
        \includegraphics[width=\linewidth]{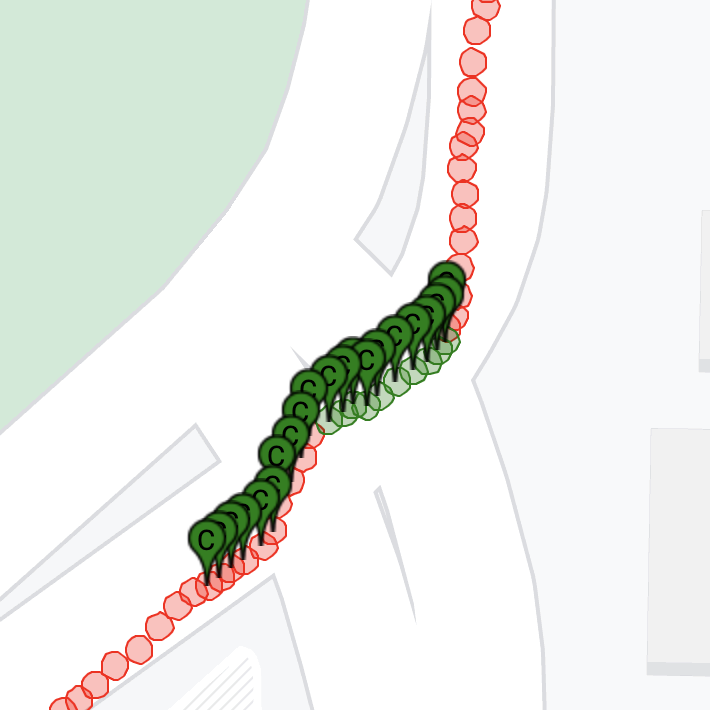}
        \caption{}
        \label{fig:subfig6}
    \end{subfigure}

    \caption{We visualized six walking traces on a map, each marked with annotated behaviors and prediction results. Red dots represent GPS coordinate points of a walking trace. Here, a true positive prediction is indicated by a green dot, and a false positive prediction by a blue dot. A green marker with ``C'' and a yellow marker with ``P'' represent actual road \textbf{C}rossings and \textbf{P}otential road crossings, respectively, as labeled by annotators. (a) The participant is not crossing, but severe GPS drift causes false alarms; (b) Significant GPS delay leads to false alarms, also called late warnings; (c) GPS drifts to the other side of the road. \name avoids initial false alarms but later issues a short-period false alarm; (d) GPS freezes while the participant is crossing. \name predicts true crossings but also issues false alarms afterward; (e) The participant shows crossing intent as labeled by annotators, and \name predicts crossing, which is also categorized as a false alarm; (f) Despite continuous GPS drift, \name successfully distinguishes between true and false crossings.}
    \label{fig:prediction_sample}
\end{figure}

% \todo{more analysis here} 
% \todo{pls revise this section, don't jump into false positive directly. what are all types of false? dont use false alarm}
We analyzed the root causes of false positive predictions and false negative predictions here. False positive predictions, also called false alarms in \name, can be triggered by a variety of factors, including GPS errors, crossing potentials, road shapes, and labeling discrepancies. We visualized each participant's walking trace on a map with the three annotated behaviors (443 complete crossings), false positive and false negative predictions highlighted. We carefully examined the visualized traces and categorized the overall \textbf{64} false alarms out of the 473 ``True'' predictions on the semi-controlled traces based on their root causes. We illustrated six walking traces for better explanation in Figure \ref{fig:prediction_sample}. 

\begin{itemize}
    \item  \textbf{GPS drifts}: Although \name can tolerate GPS errors to some extent, severe GPS drifts, as in Figure \ref{fig:subfig1} and Figure \ref{fig:subfig3}, can still mislead \name. \textbf{23} false alarms, approximately 5\% of all crossing predictions, were attributed to such drifts, where the GPS data incorrectly indicated a pedestrian's entry onto the roadway.
    
     \item \textbf{Late warnings:} \name sometimes issued a crossing alert after the pedestrian has already completed the crossing. It resulted from GPS signal delays as indicated in Figure \ref{fig:subfig2} and Figure \ref{fig:subfig4}. Late warnings account for \textbf{8} false alarms. Additionally, these GPS delays can also result in false negative predictions. 
    
     \item \textbf{Crossing potentials:} \name issued \textbf{11} ``false'' alarms attributed to participants' crossing potentials. We identified these alarms resulting from data being labeled as ``Potential Crossing'' and ``Dangerous Behavior'' without an actual crossing. During these events, participants' turning towards the road or approaching to the road's edge caused \name to issue crossing alerts, as shown in Figure \ref{fig:subfig5}. For two annotators, \name triggered alarms for 5 out of 43 and 11 out of 63 labeled potential crossing events, respectively. 

     \item \textbf{Road shape:} Of the \textbf{10} false alarms attributed to road shape, 7 occurred near a curvy road where pedestrians were heading towards the road's center. The remaining three false alarms occurred near crossings where pedestrians turned right rather than continuing straight.  

     \item \textbf{Labeling discrepancy:} \textbf{12} false alarms can be attributed to labeling discrepancies. Seven stemmed from the OSM database mistakenly identifying bike paths as roads. Four were due to annotators missing labeling two court roads. In the last false alarm, the pedestrian completed the study while standing by the road and facing it, yet neither annotator classified this as a potential crossing.

\end{itemize}
 % \todo{Fixed. what is the method for the error analysis methodology? need one or two sentences. Be more accurate and conclusive about the numbers, you can put the numbers along with the bold bullets. need some discussion about how to further reduce these errors, can add it later. When needed, add figure as example}

False negative predictions are caused by GPS delays and drifts. Although OHA heading indicates pedestrian facing towards the road, \name does not issue a crossing alert since pedestrian seems static (GPS delay, 8 cases) or far away from the road (GPS drift, 21 cases), constituting approximately 6\% of all crossing predictions.

%%% 21 cases out of what??? these numbers are small!!

%%%%%%%%%%%%%%%%%%%%%%%%%%%%%%
% show prediction results and error analysis graph

%%%%%%%%%%%%%%%%%%%%%%%%%%%%%%

\subsubsection{Model Explainability}
\begin{figure}
     \centering
         \includegraphics[width=0.6\linewidth]{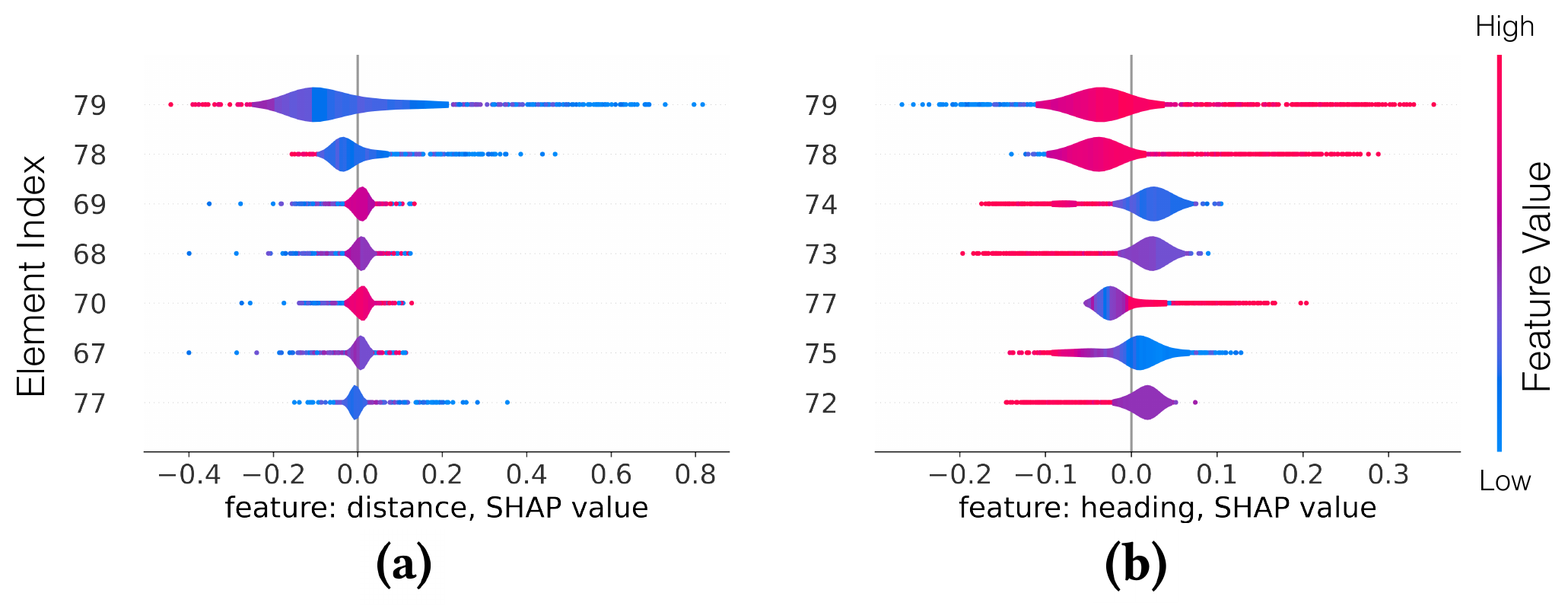}
        \caption{SHAP analyzes \name model input features: (a) distance and (b) heading. X axis represents SHAP value(impact on model output, positive means more likely to cross, negative means more likely to not cross), and Y axis shows top 7 important time indexes. }
        \label{fig:shap}
\end{figure}

We used SHAP (SHapley Additive exPlanations \cite{NIPS2017_7062}) to analyze how each feature contributes to our model's predictions. We randomly selected 1200 samples from our user study, each of length 80 elements and with two features. We displayed the top 7 most important time indexes of both input features in Figure~\ref{fig:shap}. The X-axis represents the impact on the model output, and the Y-axis denotes the element index, with 79 being the last and most important element index of the input vector.
We observed that the model attributes similar importance to heading information and coarse-grained distance information.
Furthermore, the model receives positive impact with high heading value (cosine value equal to 1 indicates the OHA heading and Road reference angle are aligned) and low distance value (a smaller distance to the road center suggests nearing the road center) on index 79. However, the model's behavior is the opposite for indexes around 70.

From the samples, we observed that the model may give non-crossing predictions when pedestrians turn away from the road, in which index 70 may have a high heading value. Additionally, the last 3 to 4 seconds of data usually have higher impacts on prediction results than the initial ones. 
% \todo{Fixed. how does the second correspond to index?}

%%%%%%%%%%%%%%%%%%%%%%%%%%%%%%

\subsubsection{ Baseline Model Comparison}\label{sec:baseline}
We trained baseline models using alternative input features for comparative analysis. Instead of using our curated features described in Section \ref{sec:feature_extract}, we utilized three input feature sets: 1) distance to road center only, 2) GPS bearing only, and 3) a combination of both distance to road center and GPS bearing. It is important to note that these features are directly derivable from GPS data and road shape information, with the distance to road center same as that in our primary model. We also chose a \textit{lookback} length of 8 seconds to facilitate a more straightforward comparison. We trained and evaluated the baseline models with same training parameters, train/test set and evaluation set. 

Baseline model with feature set one achieved 78\% precision rate and 86.8\% recall rate on the test set of daily walking traces, and 84.4\% precision and 90.7\% recall on the video-recorded walking traces. 
In terms of early detection time, our model predicts pedestrian crossings prior to reaching the road edge. In contrast, the baseline model typically identifies a true crossing, on average, \textbf{2.1 seconds after} crossing the road edge, rendering it less effective for real-world applications. Moreover, such advance in early detection time, as well as visualized walking traces in Figure \ref{fig:subfig4} and Figure \ref{fig:subfig6} demonstrate that \name can mitigate the impact of GPS inaccuracies with the assistance of fine-grained real-time heading. 

Furthermore, models trained with the second and third feature sets, both incorporating GPS bearing, exhibited significantly poorer performance. They failed to effectively learn from the data, regardless of variations in model architecture or adjustments in the \textit{lookback} duration. We hypothesize that this inadequacy stems from the inherent GPS inaccuracies. Such inaccuracies might lead to false indications of pedestrian movement towards the road, even when no actual road crossing occurs. Consequently, the inclusion of such features, which are not robust or reliable, misleads the model.

\subsection{\textbf{Q3. System Overhead}}
% note: when GPS accuracy is very low, larger than 10 meters, choose to not report true road crossing alert, 
% user evaluation, real-world
% road condition and different user's walking habit 
We evaluate \name's system overhead in both active mode and idle mode. 
% We also evaluate the wireless delay and success rate of our alert broadcasting method.\smallskip
% IMWUT2024

\begin{table*}
\centering
\resizebox{\textwidth}{!}{
\begin{tabular}{@{}ccccccccccc@{}}
\toprule
\textbf{Device} & \textbf{CPU} & \textbf{RAM} & \textbf{DB Query} & \textbf{OHA} & \textbf{Inference} & \textbf{Execution} & \textbf{CPU Usage \%} & \textbf{MEM Usage \%} & \textbf{Consumed} \\ 
\textbf{specs}& \textbf{cores}&\textbf{(GB)} & \textbf{(ms)} & \textbf{(ms)} & \textbf{(ms)} & \textbf{(ms)} & \textbf{(High/Low/Avg)} & \textbf{(High/Low/Avg)} & \textbf{Power(mAh) } \\
\midrule
Pixel 3, Android 11 & 8 & 4 & 63.91 & 0.068 & 9.93 & 10.24 & 100.0/87.3/92.3 & 4.5/3.8/4.25 & 289.2 \\
Pixel 6, Android 12 & 8 & 8 & 32.92 & 0.054 & 7.01 & 7.63 & 120.0/73.0/103.0 & 4.0/2.5/3.03 & 421.1 \\
Pixel 7, Android 13 & 8 & 8 & 58.04 & 0.088 & 7.72 & 8.27 & 101.0/89.0/95.0 & 3.1/2.7/2.93 & 346.4 \\
\bottomrule 
\end{tabular}
}
\caption{Device specs, execution time and resource usage when running \name in active mode.}
\label{table:sys_overhead}
\end{table*}

\noindent \textbf{Measurement Setup:}
% \begin{itemize}
%     \item task completion time
%     \item power consumption
%     \item CPU usage history, high low utility rate
% \end{itemize}
% Our setup for evaluating the system overhead of \name comprises three Pixel 6 devices with Octa-core (2x2.80 GHz Cortex-X1, 2x2.25 GHz Cortex-A76, and 4x1.80 GHz Cortex-A55) and 8Gb RAM, operating on both Android 11 and Android 12. 
Our setup for evaluating the system overhead of \name includes two Pixel 3 devices, three Pixel 6 devices, and one Pixel 7 device. We charged these smartphones to full battery capacity and had \name deployed and properly configured for each smartphone.

We first measured the program execution time. \name incorporates various functions, including database query, feature extraction, and model inference. We utilized the Android system API to record the execution time of each function in nanoseconds.
We report the average execution time of each function when running \name for 30 minutes continuously.
We used the Android \texttt{BATTERY PROPERTY CHARGE COUNTER} to retrieve the current battery capacity in mAh during the run time. To minimize the influence of other apps, only our app and the PostgreSQL database were active during the measurement of system overhead.

% \todo{update results, add pixel 5,7 results, show as table}
\noindent \textbf{Overhead Results:}
We present the system overhead results of active mode in Table~\ref{table:sys_overhead}. The average execution time for each prediction (OHA, inference and others) is around 10 ms. Noted that database queries were conducted in a separate thread and were executed once per second.
In idle mode, we observed a significant drop in average resource utility and power consumption. Specifically, the average CPU utility rate decreased to approximately 0.56\%, and the average RAM usage fell to about 2.4\%. Power consumption, when continuously running \name for 30 minutes, decreased to approximately 95 mAh across all devices.

\section{Related Work}

Pedestrian road crossing prediction is a popular research topic due to its potential to protect the safety of road users. Existing research in this area mainly uses cameras or LiDARs on vehicles and infrastructure~\cite{schulz2015controlled,keller2013will,mogelmose2015trajectory,minguez2018pedestrian, wu2018novel}. For completeness, we also discuss infrastructure-free approaches that apply physical modeling or machine learning methods to predict pedestrian mobility. The latter approaches do not consider road crossing prediction per se, yet they can help such a system when combined with road map data. 

\subsection{Infrastructure and Vehicle-based Approaches}
Infrastructure-based approaches utilize sensors, such as LiDARs and cameras, to first recognize pedestrians in a scene and then track their poses and positions relative to roads~\cite{sighencea2021review, schulz2015controlled}. Keller et al. estimate fine-grained pedestrian positions and poses from a camera. Then, they classify pedestrian movements, including standing, stopping, and crossing~\cite{keller2013will}. Mogelmose et al. fuse pedestrian's locations from the camera, vehicle's location, and OpenStreetMap data to decide whether a pedestrian is within a danger zone~\cite{mogelmose2015trajectory}. Minguez et al. provide accurate pedestrian activity recognition by estimating pedestrian gait from joint data and using a dynamic model to predict pedestrian's movements~\cite{minguez2018pedestrian}.

Besides detection accuracy, time-to-crossing, is another key factor in evaluating such pedestrian crossing detection systems. Zhang et al.~\cite{zhang2021pedestrian} predict the crossing intent at intersections by estimating pedestrian poses captured by CCTV cameras. Their system achieves its peak performance, as measured by the F1 score when forecasting crossing intentions approximately one second prior to the pedestrian initiating the crossing.
Jeong et al.\cite{jeong2016early} and Xu et al.\cite{xu2011detection} classify sudden pedestrian crossings into three danger levels—warning, caution, and normal—based on the percentage of overlap between the pedestrian's body and the road reference line. Fang et al.~\cite{fang2019intention} evaluated the performance of recent advancements in vision-based human pose estimation for identifying pedestrian crossing intentions. They found that in scenarios where the subject continued walking to cross the road, the probability of crossing increased to 0.8 as early as 15 frames (0.5 s) ahead of the crossing events.

For pedestrians who are occluded behind objects such as parked cars, these systems cannot predict pedestrians' behavior without capturing the pedestrians through cameras. Palffy et al.~\cite{palffy2019occlusion} proposed an occlusion-aware model utilizing cameras and LiDAR reflections to help detect pedestrians earlier. Their system can capture pedestrians 0.3 seconds earlier than a camera-based detector, which detects the pedestrians after they appear in sight. 
 
% reviewed past papers about pedestrian motion prediction and reported the pitfalls in those approaches. They
Scholler et al.~\cite{scholler2020constant} claimed environmental priors and long pedestrian motion history might negatively impact the generalization of these models. 
Other researchers track human gestures through Wi-Fi \cite{jiang2020towards,zheng2019zero, fang2020fusing} or millimeter-wave \cite{kong2022m3track,zhang2022synthesized}, which is not applicable in outdoor environments. 
% Alternative localization options, like those based on 5G~\cite{5gposition}, promise submeter accuracy, but they are not widely deployed, require special hardware, and raise privacy concerns.

Compared to these approaches, \name does not require line of sight, infrastructure, or special hardware since it predicts pedestrian behavior utilizing smartphone sensors. Table \ref{tab:time_to_crossing} shows how \name compares in performance to prior approaches. The main takeaway from the table is that \name provides superior performance to camera-based approaches with an acceptable prediction delay.

\begin{table}[ht]
\centering
\caption{Comparison of Time-to-Crossing for pedestrian crossing detection. LOS represents Line-of-Sight, and NLOS represents Non-line-of-sight. Time-to-crossing refers to the time before crossing the road edges at which the prediction occurs. Higher time-to-crossing indicates earlier detection. Probability refers to its confidence in determining whether a pedestrian is going to cross.}
\label{tab:time_to_crossing}
\begin{tabular}{@{}lllllll@{}}
\toprule
Author(s)       & Conditions & Sensors Used         & Time-to-Crossing (s)  & Precision & Recall &  Probability \\ \midrule
Xu et al.~\cite{xu2011detection}       & LOS        & Camera (vehicle)     & 0 & 0.52      & -   & -   \\
Keller et al.~\cite{keller2013will}   & LOS        & Camera (vehicle)     & Up to 0.77s           & -         & -  & -    \\
Jeong et al.~\cite{jeong2016early}    & LOS        & Far-infrared camera     & 0 & 0.85      & -   & -   \\
Palffy et al.~\cite{palffy2019occlusion}   & LOS, NLOS  & Camera, Radar     & 0.3                   & -       & -   & 0.8   \\
Fang et al.~\cite{fang2019intention}    & LOS        & Camera (vehicle)               & 0.43 (13 frames)      & -       & -   & 0.8   \\
Zhang et al.~\cite{zhang2021pedestrian}    & LOS        & Camera (road-side)   & 1                     & 0.859     & 0.818  & - \\
\textbf{Ours: \name }       & LOS, NLOS  & IMU, GPS             & 0.39                  & 0.869     & 0.936  & - \\ \bottomrule
\end{tabular}
\end{table}

\subsection{Infrastructure-free Approaches}

The other category of approaches uses IMU sensors from the pedestrian to predict their mobility without relying on external infrastructure.

\subsubsection{Kinematics-based Approaches}
These approaches, such as dead-reckoning, apply motion dynamics on IMU sensors and location data to model pedestrian movements~\cite{wu2019survey}. Feigl et al.~\cite{feigl2019bidirectional} utilize IMU sensor and deep learning models to predict the pedestrian's velocity. Kuang et al.~\cite{kuang2018robust} developed a pedestrian trajectory prediction approach that performs heading estimation, position estimation, and step detection using smartphone IMU sensors. 
Fan et al.~\cite{fan2019improved} further improved the positioning accuracy by adding an adaptive Kalman filter to heading estimation. Other dead-reckoning approaches utilize a step-and-heading method involving step counting, step stride estimation, and heading estimation. Edel et al. utilize Bi-LSTM and RNN models to achieve this task from IMU sensors on the phone~\cite{edel2015advanced}. Xing et al. use specific sensor hardware installed on the foot to predict step length~\cite{xing2017pedestrian}. However, IMU sensors are prone to sensor drifts in the long term, and these methods did not demonstrate their performance in such situations when pedestrians keep walking outdoors. Map-matching techniques, such as open source routing machine (OSRM)~\cite{project-osrm}, can match GPS coordinates of walking traces to the nearest road or pedestrian paths. However, these techniques may fail if GPS continuously drifts, potentially causing incorrect mapping to the opposite side of the road.

Related to the above, some approaches track the direction of the pedestrian using smartphone sensors. Roy et al. proposed ``WalkCompass'' to estimate the user's walking direction from a smartphone~\cite{roy2014smartphone}. In particular, they compute the motion vector using the accelerometer and gyroscope, calibrate the compass sensor, and determine the user's walking direction. However, the direction accuracy drops significantly if the user swings their smartphone or changes its position. Zhou et al. proposed a smartphone orientation estimation method, $A^3$, using the gyroscope, accelerometer, and compass sensor on commodity smartphones~\cite{zhou2014use}. Shen~\cite{shen2018closing} further improved the estimation accuracy by introducing the magnetometer sensor for calibration and proposed MUSE. As the state-of-the-art methods for estimating smartphone orientation, $A^3$ and MUSE yield high accuracy when environmental magnetic field interference is small.

These methods are limited because smartphone orientation does not provide a direct indication of the pedestrian's actual heading, and they are highly volatile in response to common human behaviors, such as the swinging of the smartphone or variations in the pedestrian's gait. \name addresses this problem by introducing a simple but practical algorithm to estimate the pedestrian's actual heading. More importantly, it introduces a new model to predict crossing based on recent mobility features.

\subsubsection{Machine learning-based Approaches}
Machine learning-based methods reconstruct the pedestrian's trajectory through training models on past IMU sensor data. Chen et al. proposed IONet to learn pedestrian trajectory from IMU measurement data through Recurrent Neural Networks and directly reconstruct the pedestrian's trajectory~\cite{chen2018ionet}. Although IONet shows better accuracy than traditional methods, its errors accumulate rapidly as the user continues moving. DeepIT significantly improves trajectory estimation accuracy over IONet with the assistance of earbud sensor data~\cite{gong2021robust}. With sensor data from both a smartphone and an earbud, DeepIT estimated the reliability of sensor data and compensates for the angle drift caused by sensor inaccuracies. Both IONet and DeepIT achieve relatively high precision in estimating the spatial translation of a user's location while less accurate in estimating directional orientation.

\section{Limitations}
\label{sec:reduce_fl}
We describe the limitations in the evaluation and design of \name. 

% \noindent 
% \textbf{Orientation Estimation:} \name selects magnetic field constraints through extensive analysis in our daily walking traces. However, this constraint may degrade performance in other places due to geomagnetic field distributions. \smallskip

\noindent 
\textbf{OHA Algorithm:} \label{sec:oha_limit}
During the initialization of $R_{LP}$, we approximate the pedestrian heading with GPS bearing. However, the OHA algorithm might produce distorted headings due to erroneous GPS signals encountered during this phase. OHA's effectiveness also diminishes if the GPS heading is persistently inaccurate or if pedestrians intentionally position their phones in locations with the same roll-pitch pairs.
Additionally, two relative orientation matrix could share identical $\theta$ and $\psi$ angles, leading to OHA heading errors. However, given a large number of possible $(r,p)$ pairs, specifically 16,200 for quantization factor as 2, the collision likelihood is minimal. To mitigate this, \name restarts the OHA algorithm whenever a pedestrian begins a new walk. 

\noindent
\textbf{Time-to-crossing:}
\name achieves performance comparable to camera-based pedestrian crossing prediction techniques, while maintaining a relatively higher precision and recall rate. Additionally, \name issues alerts before a pedestrian crosses the road edge, providing a one to two-second window before the pedestrian reaches the potential collision point. However, considering the wireless broadcasting delay and the driver's response time, nearby drivers might not have sufficient time to maneuver. Assuming the vehicle travels at speeds between 25 and 40 mph on flat, dry urban roads, and considering a driver response time of 1 second, the vehicle can come to a full stop within a stopping distance ranging from 20 meters to 41.2 meters~\cite{szyk2023stopping}. Therefore, our system can effectively alert nearby drivers out of this stopping distance. Meanwhile, our future work will focus on personalizing the prediction model to further increase time-to-crossing.

\noindent 
\textbf{False Alarms:} 
GPS delay causes both false alarms and missed crossing alerts since it postpones \name's prediction on when the pedestrian should enter the road range. Future work will focus on monitoring pedestrians' motion toward the road and validating if the GPS signal has been frozen. Adding reliability metrics on coarse-grained distance and heading features should further eliminate false alarms.

\section{Conclusion}

% This paper introduces \name, an infrastructure-free and ready-to-use system that predicts pedestrian road crossing using commodity smartphones. Through extensive data collection in daily life and road-crossing behavior analysis, we demonstrate \name's capability to identify road-crossing behaviors before a pedestrian enters the road range. \name addresses multiple challenges, including GPS errors, IMU drifts, and diversity in pedestrian mobility, while minimizing the inevitable false alarms inherent in developing such a system. Our future research includes integrating the networking component for the dissemination of crossing alerts, designing personalized models that consider individual pedestrian behaviors, and designing driver alert mechanisms to balance pedestrian safety with system utility for drivers. 

This paper proposes a new heading tracking algorithm, OHA, that accurately tracks pedestrian headings using smartphones, regardless of the smartphone's orientation. We evaluated OHA across nine distinct scenarios, where it achieved consistently low heading errors over extended periods compared to other baseline techniques, even in the most challenging scenario. Additionally, we employed the OHA heading in predicting pedestrian crossing behaviors to enhance road user safety in real-world environments. We introduce \name, an infrastructure-free and ready-to-use system that predicts pedestrian road crossings using commodity smartphones. Through extensive data collection in daily life and semi-controlled environments, \name demonstrates that precise heading enables the early detection of pedestrian crossing behaviors, issuing crossing alerts on average 0.35 seconds before pedestrians enter the road range.

\bibliographystyle{ACM-Reference-Format}
\bibliography{sigproc}

\appendix

\end{document}